\documentclass[twocolumn,showpacs,preprintnumbers,superscriptaddress,amsmath,floatfix,amssymb,secnumarabic]{revtex4}
\usepackage{graphicx}
\usepackage[colorlinks=false]{hyperref}

\newcommand{\comment}[1]{}

\newcommand{\fm}{\, {\rm fm}}
\newcommand{\mev}{\, {\rm MeV}}

\newcommand{\gevq}{\, {\rm GeV^2}}

\newcommand{\lr}[1]{ \left( #1 \right) }
\newcommand{\lrs}[1]{ \left[ #1 \right] }

\newcommand{\vev}[1]{ \langle \, #1 \, \rangle }

\newcommand{\sign}{ {\rm sign} \,  }

\begin{document}
\sloppy

\title{Deconfinement transition in two-flavour lattice QCD with dynamical overlap fermions in an external magnetic field.}

\author{V.~G.~Bornyakov}
\affiliation{Institute for High Energy Physics, 142281, Protvino, Russia}
\affiliation{Institute of Theoretical and Experimental Physics, 117259 Moscow, Russia}
\affiliation{Far Eastern Federal University, Sukhanova Street 8, Vladivostok 690950, Russia}

\author{P.~V.~Buividovich}
\affiliation{Institut f\"{u}r Theoretische Physik, Universit\"{a}t Regensburg, D-93040 Regensburg, Germany}

\author{N.~Cundy}
\affiliation{Lattice Gauge Theory Research Center, FPRD, and CTP  \\
Department of Physics and Astronomy, Seoul National University,Seoul, 151-747, South Korea}

\author{O.~A.~Kochetkov}
\email{Oleg.Kochetkov@physik.uni-regensburg.de}
\affiliation{Institut f\"{u}r Theoretische Physik, Universit\"{a}t Regensburg, D-93040 Regensburg, Germany}
\affiliation{Institute of Theoretical and Experimental Physics, 117259 Moscow, Russia}

\author{A.~Sch\"{a}fer}
\affiliation{Institut f\"{u}r Theoretische Physik, Universit\"{a}t Regensburg, D-93040 Regensburg, Germany}

\date{August 15th, 2014}

\begin{abstract}
 We study the influence of an external magnetic field on the deconfinement transition in two-flavour lattice QCD with physical quark charges. We use dynamical overlap fermions without any approximation such as fixed topology and perform simulations on a $16^3 \times 6$ lattice and at a pion mass around $500 \mev$. The pion mass  (as well as the lattice spacing)  was determined in
independent runs on $12^3 \times 24$ lattices. We consider two temperatures, one of which is close to the deconfinement transition and one which is above. Within our limited statistics the dependence of the Polyakov loop and chiral condensate on the magnetic field supports the ``inverse magnetic catalysis'' scenario in which the transition temperature decreases as the field strength grows for temperature not to far above the critical temperature.
\end{abstract}
\maketitle

\section{Introduction}
\label{sec:intro}

 The influence of strong magnetic fields on the quark-gluon plasma is at present a subject of intensive research. Magnetic fields of order $0.1 - 1.0 \gevq$ can be produced for a short period of time in heavy ion collisions in RHIC and LHC \cite{Skokov:09, Skokov:13}. Such a magnetic field strength is comparable to the pion mass squared and hence can lead to novel effects at the hadronic scale, e.g. related to strong $\mathcal{CP}$ violation \cite{Kharzeev:08:1}. These effects are the topic of controversial ongoing discussions. In recent years it became clear that in order to incorporate the effects of local $\mathcal{CP}$ violation the description of the quark-gluon plasma by means of conventional second-order relativistic viscous hydrodynamics (for a nice review see \cite{Romatschke:09:1}) should be amended by including the transport coefficients and chemical potentials which are odd under $\mathcal{CP}$ \cite{Son:09:1, Sadofyev:10:1, Sadofyev:10:2, Zakharov:12:1, Jensen:12:1, Jensen:12:2, Banerjee:12:1}. Such an ``anomalous hydrodynamics'' requires an equation of state with new ``chiral'' chemical potentials and the external magnetic field. One also needs the values of the transport coefficients as input which should be provided, for example, by lattice simulations.

 Therefore one of the important ingredients in the self-consistent hydrodynamical description of the influence of external magnetic fields on heavy-ion collisions is the equation of state of strongly interacting matter at non-zero magnetic field and/or non-zero chiral chemical potential. While a consistent construction of the overlap Dirac operator with non-zero chiral chemical potential was found only recently \cite{Buividovich:13:6, Buividovich:13:8}, the introduction of external magnetic field in the existing simulation algorithms for overlap fermions is straightforward. Thus with rather easy modification of existing codes one can answer the important question of how super-strong magnetic fields influence the temperature of the deconfinement phase transition.

 Before detailed lattice studies were performed in \cite{D'Elia:10,Endrodi:12:jhep,Endrodi:13:jhep,Endrodi:12:prd,Muller-Preussker:13}, the general belief was that ``magnetic catalysis'', that is, the growth of the chiral condensate and thus of the transition temperature with magnetic field strength, is a solid QCD prediction \cite{Smilga:97:1}. A large number of effective models, including chiral perturbation theory \cite{Smilga:97:1, Cohen:07, Agasian:01, Andersen:12:1}, functional renormalization group \cite{Andersen:12:2,Andersen:13}, the Gross-Neveu model \cite{Scherer:12,Sato:98}, the Sakai-Sugimoto model at zero chemical potential \cite{Johnson:08}, the linear sigma model \cite{Fraga:08} and the extended Polyakov-Nambu-Jona-Lasinio model \cite{Ruggieri:2011, Ruggieri:10, Kashiwa:11}, agree with the ``magnetic catalysis'' scenario. At zero temperature this behaviour can be related to the $\beta$ function of scalar QED and is thus not a property of QCD as such \cite{Endrodi:2013cs}. At higher temperatures the dependence of the condensate on the magnetic field is determined by QCD interactions, but one can still expect that for not very large temperatures the growth of the condensate with magnetic field will persist. In contrast, lattice simulations \cite{D'Elia:10,Endrodi:12:jhep,Endrodi:12:prd,Muller-Preussker:13} have demonstrated that the QCD response at sufficiently high temperatures is instead very sensitive to, e.g., the quark mass. Explicit two-loop calculation of the magnetic correction to the pressure \cite{Fraga:13plb} as well as calculations within field-correlators framework \cite{Simonov:14} also supports this observation .

 The first lattice study of the deconfinement transition in the presence of an external magnetic field \cite{D'Elia:10} was performed with rooted staggered fermions at larger than physical pion mass. It also indicated that a magnetic field increases the deconfinement temperature. In \cite{Endrodi:12:jhep,Endrodi:12:prd} a similar study was made with several different values of the $u$- and $d$-quark masses, ranging from the strange quark mass down to their physical values. It was found that for sufficiently light quarks (in other words, for sufficiently small pion mass) the deconfinement temperature starts decreasing with magnetic field. This scenario is usually referred to as ``inverse magnetic catalysis''. In addition, in a recent work \cite{Muller-Preussker:13} of the Berlin group, some signatures of inverse magnetic catalysis were found at high temperatures and magnetic fields $B<0.8 \gevq$ for $N_f=4$ flavours of staggered fermions (without rooting) and close to physical pion mass in two-color lattice QCD. An interesting phenomenological picture of inverse magnetic catalysis resulting from a delicate interplay of low-lying Dirac eigenvalues and the Polyakov loop was proposed in \cite{Endrodi:13:new}.

 Inverse magnetic catalysis can be also successfully reproduced in the large-$N_c$ limit of QCD \cite{Fraga:13prd} as well as in some effective models of QCD, such as the antipodal Sakai-Sugimoto model \cite{Ballon-Bayona:13}, the bag model \cite{Fraga:12} and in the Polyakov-Nambu-Jona-Lasinio models with the coupling constant which depends on the magnetic field \cite{Ferreira:14,Ferreira:14prd}. An interesting mechanism which can explain the inverse magnetic catalysis is the strong enhancement of the fluctuations of axial charge density $j^A_0 = \bar{\psi} \gamma_0 \gamma_5 \psi$ in the presence of external magnetic field \cite{Chao:13,Yu:14:1}. Since at nonzero axial charge density the Fermi levels for left- and right-handed fermions shift in the opposite directions, the formation of the bound states $\bar{\psi}_R \psi_L$ and $\bar{\psi}_L \psi_R$ which constitute the chiral condensate is disfavoured. Thus nonzero axial charge density tends to decrease the condensate and thus lower the critical temperature of the restoration of chiral symmetry \cite{Ruggieri:11:1, Gatto:12:1, Nedelin:11:1}. A crucial ingredient in this mechanism are the interactions in the axial-vector iso-scalar channel, which become repulsive at $T \gtrsim T_c$ due to instanton-anti-instanton molecule pairing \cite{Shuryak:94:1, Shuryak:94:2}.

 The results of these works at least qualitatively agree with the lattice data. We can conclude that while some generic features of the QCD phase transition can be successfully reproduced by almost all effective models on the market, the decrease of the deconfinement temperature in an external magnetic field is quite a specific prediction. Thus, inverse magnetic catalysis turns out to be extremely sensitive to the balance between quark and gluon degrees of freedom and therefore has a potential to exclude a large number of effective models which were proposed to describe high-energy heavy-ion collisions.

 To make full use of this potential it is crucial to reduce not only the statistical but also the systematic uncertainties of lattice simulations. In particular, we notice that almost all previous lattice studies of the influence of a magnetic field on QCD with dynamical quarks were done with rooted staggered fermions. A question of whether or not rooted staggered fermions correctly reproduce the instanton-mediated interactions between quarks (t'Hooft vertex) has been a subject of intense debates in recent years \cite{Creutz:07:1, Creutz:08:1, Creutz:08:2, Bernard:08:1, Bernard:08:2, Kronfeld:11:1}. But exactly such type of interactions (with certain finite-temperature modifications \cite{Shuryak:94:1, Shuryak:94:2}) is crucial for the recently proposed scenario \cite{Chao:13, Yu:14:1} of the inverse magnetic catalysis. While rooted staggered fermions might still reproduce the required axial-vector iso-scalar vertex in the continuum limit \cite{Bernard:08:1, Bernard:08:2, Kronfeld:11:1}, one can expect that at finite lattice spacing the interactions between truly chiral lattice fermions which are mediated by topological objects will be closer to the ones in the continuum theory.

 Also, for applications to heavy-ion collisions the phase diagram of magnetized QCD matter is mostly interesting in conjunction with the values of the anomalous transport coefficients (most notably the chiral magnetic \cite{Kharzeev:09:1} and the chiral separation \cite{Son:06:2, Metlitski:05:1} conductivities), which can get nontrivial radiative corrections in interacting theories \cite{Jensen:13:1,Buividovich:13:8,Gursoy:14:1}. Therefore one would like to obtain both the equation of state and the values of transport coefficients using a single lattice action. Recent works \cite{Buividovich:13:6, Buividovich:13:8} by one of the authors showed that the correct lattice implementations of chiral symmetry and the corresponding $U\lr{1}_A$ axial anomaly are essential for the correct definition of anomalous transport on the lattice. Since the tastes of staggered fermions have opposite chiralities, it is not clear whether it is possible to reproduce anomalous transport with staggered fermions. For instance it might turn out that the Chiral Magnetic and the Chiral Separation effects simply cancel between the tastes. Also for this reason it is important to perform lattice studies of the properties of strongly interacting chiral matter in external magnetic fields using truly chiral lattice fermions, for which anomalous transport coefficients are well-defined \cite{Buividovich:13:8}.

 As a first step towards the first principle lattice studies of the equation of state and the transport properties of chiral QCD in an external magnetic field, in this work we study the deconfinement transition in $N_f = 2$ lattice QCD with dynamical overlap fermions, using algorithms developed in \cite{Arnold:2003sx,Cundy:2004pza,Cundy:06,Cundy:09:1,Cundy:09:2,Cundy:11:2}. We consider the dependence of the chiral condensate and the Polyakov loop on the magnetic field and find that it supports the inverse magnetic catalysis scenario. We also consider the fluctuations of the topological charge, but within our statistics we cannot reach a definite conclusion on how the magnetic field changes the topological susceptibility.

 Since the magnetic field does not affect the renormalization of the physical observables which we analyse \cite{Bali:13:1, Bali:14:1}, we base our conclusions on the behaviour of non-renormalized physical observables in an external magnetic field of varying strength at fixed lattice spacing. Due to very large numerical cost of dynamical overlap fermions our statistics is still limited, however, even at the present level of accuracy we feel that  at least some qualitative statements can be made.

 This work pursues several goals: first, we want to cross-check whether the shift of the deconfinement phase transition observed with staggered fermions in \cite{Endrodi:12:jhep} is reproduced with truly chiral lattice fermions. This is important in order to quantify the predictive power of simulations both with staggered and chiral fermions and thus sharpen this tool for the falsification of models. Such comparison is also necessary to understand which bare quark mass in the overlap operator is already small enough to reproduce the inverse magnetic catalysis. Second, we test the simulation algorithms developed in \cite{Arnold:2003sx, Cundy:2004pza, Cundy:06, Cundy:09:1, Cundy:09:2, Cundy:11:2} for a lattice of reasonable size and in a physical setup for which the use of chiral fermions might be crucial. We thus demonstrate that with modern algorithms and computer resources fully dynamical simulations with overlap fermions are possible without introducing systematic errors by resorting to fixed topology or other approximations. Finally, we generate a set of gauge field configurations for further measurements of anomalous transport coefficients or other interesting properties of magnetized chiral matter near the deconfinement transition. It should be stressed that the detailed study of the order of phase transition and of its precise position is not the aim of the present paper. We plan to publish a detailed study of the deconfinement transition at zero magnetic field elsewhere.

\section{Numerical setup}
\label{sec:numerical_setup}

 We consider lattice QCD with $N_f = 1+1$ quark flavours, which have equal masses but different electric charges, $-e/3$ and $2e/3$. We use the massive overlap Dirac operator,
\begin{eqnarray}
\label{overlap_definition}
  D\lrs{\mu} = 1 + \mu/2 + \gamma_5 \lr{1 - \mu/2} \sign\lr{K} ,
\end{eqnarray}
where $K = \gamma_5 \lr{D_W - \rho}$ and $D_W$ is the Wilson-Dirac operator with one level of over-improved stout smearing \cite{Moran:2008ra,Morningstar:2003gk}. The bare quark mass is $m_q = \mu/((1-\mu)\rho) = 0.087$ (in lattice units) with $\rho=1.368, \mu=0.106$.

 In order to ensure that lattice gauge fields are sufficiently smooth, we use the tadpole improved L\"{u}scher-Weisz gauge action \cite{TILW, TILW2, TILW4, TILW5}. A special Hybrid Monte-Carlo (HMC) algorithm for overlap fermions which increases the topological tunnelling rate was used \cite{Cundy:11:2, Cundy:06, Cundy:09:1, Cundy:09:2}. Our simulations have been carried out on $16^3 \times 6$ lattices. The pion mass was calculated from the axial vector correlator $\langle \bar\psi \gamma_5 \gamma_0 \psi(x) \bar\psi \gamma_5 \gamma_0 \psi(y)\rangle$ and is around $500\mev$. The pion mass and lattice spacing were determined using independent runs on $12^3\times 24$ lattices at zero magnetic field (we do not consider here an interesting question of the pion mass dependence on the magnetic field \cite{Luschevskaya:12}).

 The temperature $T = 1/(N_t a)$ is changed by varying the inverse gauge coupling $\beta$ and thus the lattice spacing $a$. The latter was determined using the Sommer parameter \cite{Sommer:1993ce} $r_0 = 0.49 \fm$. We have performed measurements at $\beta=7.5$, which corresponds to $a=0.15\fm$ and $T=220\mev$ and at $\beta=8.3$, for which $a=0.12\fm$ and $T=280\mev$. Although the exact value of the deconfinement temperature is not known, we assume that the temperature $T=220\mev$ is close to it and that the temperature $T=280\mev$ already corresponds to the deconfinement regime. There are several arguments in favor of this assumption. First, the expectation value of the Polyakov loop at $T=220\mev$ is significantly smaller than at $T=280\mev$ (see Fig.~\ref{fig:ploop_average}) and the chiral condensate at $T=220\mev$ is significantly larger than at $T=280\mev$ (see Fig.~\ref{fig:condensate_average}).

 Second, we have also considered the distributions of the low-lying eigenvalues $\lambda$ of the projected massless Dirac operator
\begin{eqnarray}
\label{massless_dirac}
 \tilde{D}_0 = \frac{2 \rho D_0}{2 - D_0}, \quad D_0 = 1 + \gamma_5 \sign\lr{K}
\end{eqnarray}
at temperatures $T=220 \mev$ ($\beta = 7.5$) and $T=280 \mev$ ($\beta = 8.3$) as well as at intermediate values $\beta = 7.7$, $\beta = 7.9$ and $\beta = 8.1$ which correspond to some intermediate values of temperature $220 \mev < T\lr{\beta = 7.7} < T\lr{\beta = 7.9} < T\lr{\beta = 8.1} < 280 \mev$. Since for these intermediate values of $\beta$ we have not measured the lattice spacing, the exact values of these intermediate temperatures are not yet known. The factor of $\rho$ in (\ref{massless_dirac}) ensures that the eigenvalues of $\tilde{D}_0$ correspond to the eigenvalues of the continuum Dirac operator at sufficiently small lattice spacing.

 The eigenvalues $\lambda$ of $\tilde{D}_0$ are purely imaginary and are related to the chiral condensate on the lattice exactly in the same way as in the continuum theory:
\begin{eqnarray}
\label{lattice_condensate}
 \Sigma = \sum\limits_i \frac{1}{m_q + \lambda_i} = \sum\limits_{\lambda_i > 0} \frac{2 m_q}{m_q^2 + |\lambda_i|^2},
\end{eqnarray}
where $\Sigma = \frac{1}{V} \frac{\partial}{\partial \, m_q} \mathcal{Z}\lr{m_q}$ and $\mathcal{Z}\lr{m_q}$ is the lattice partition function with the Dirac operator (\ref{overlap_definition}). By virtue of the relation (\ref{lattice_condensate}), which implies that the condensate is mostly saturated by Dirac eigenmodes with $|\lambda_i| \lesssim m_q$, effective restoration of chiral symmetry should result in a significant widening of this gap.

\begin{figure*}[pH]
\includegraphics[angle=-90,width=9cm]{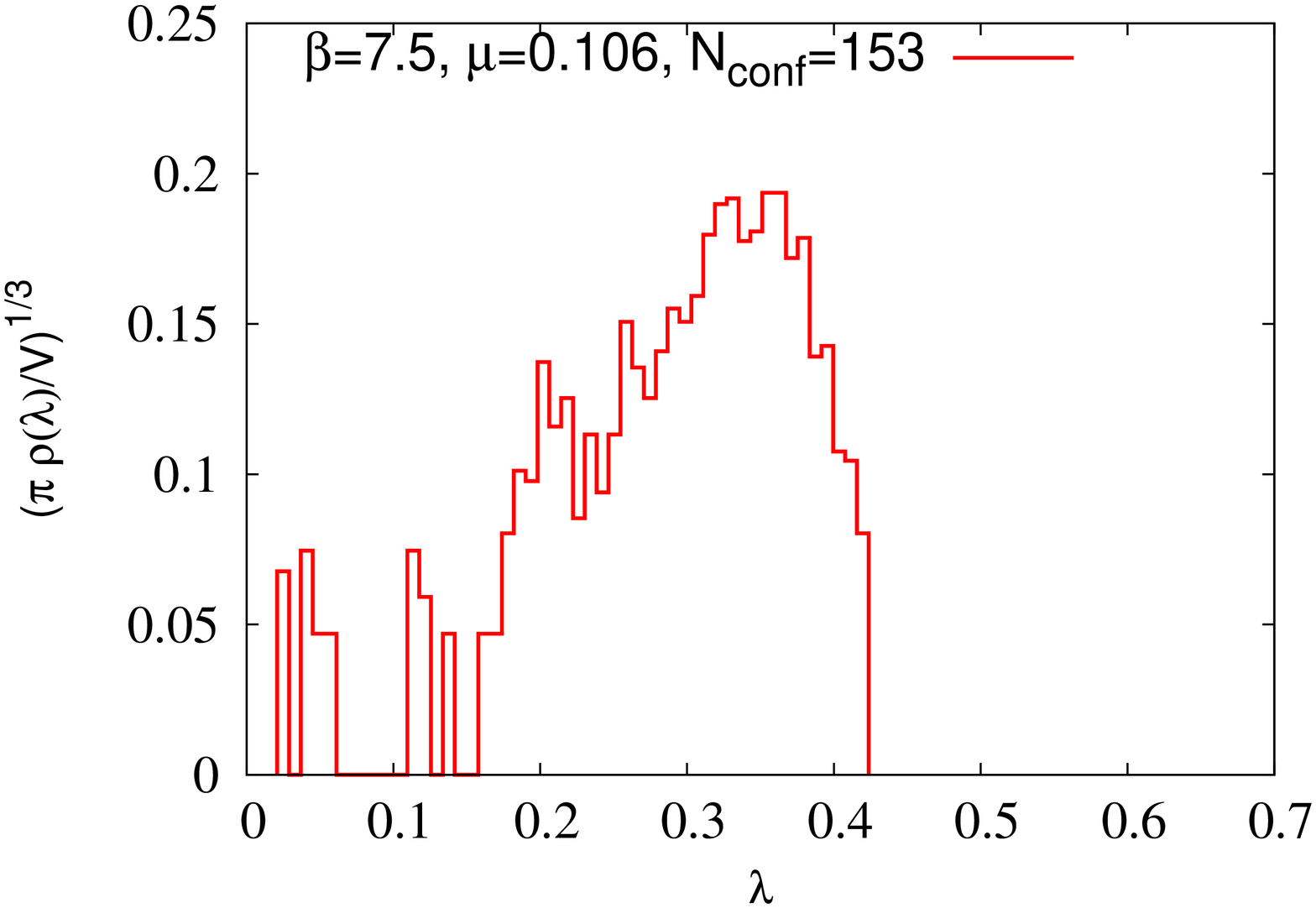}\includegraphics[angle=-90,width=9cm]{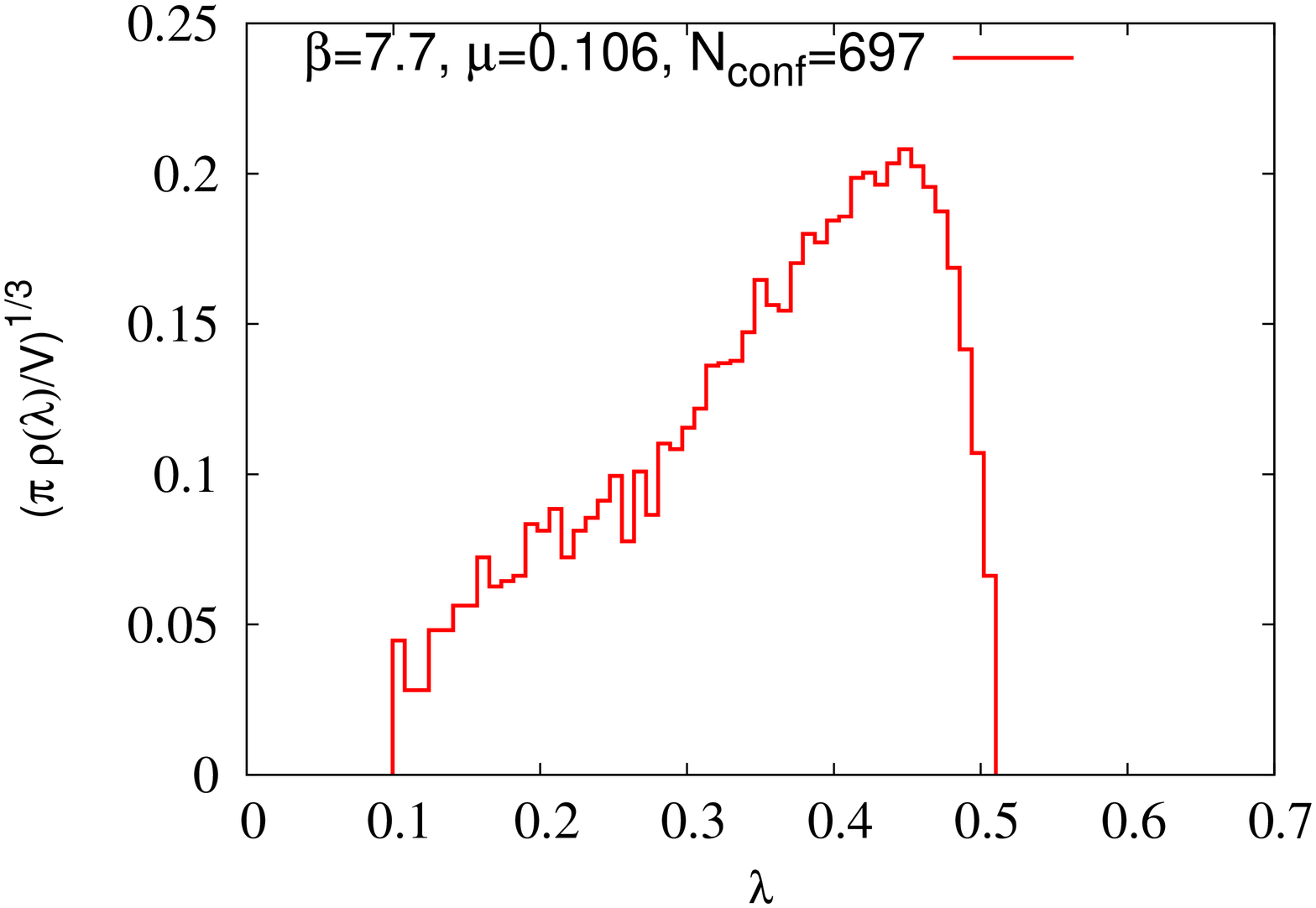}\\
\includegraphics[angle=-90,width=9cm]{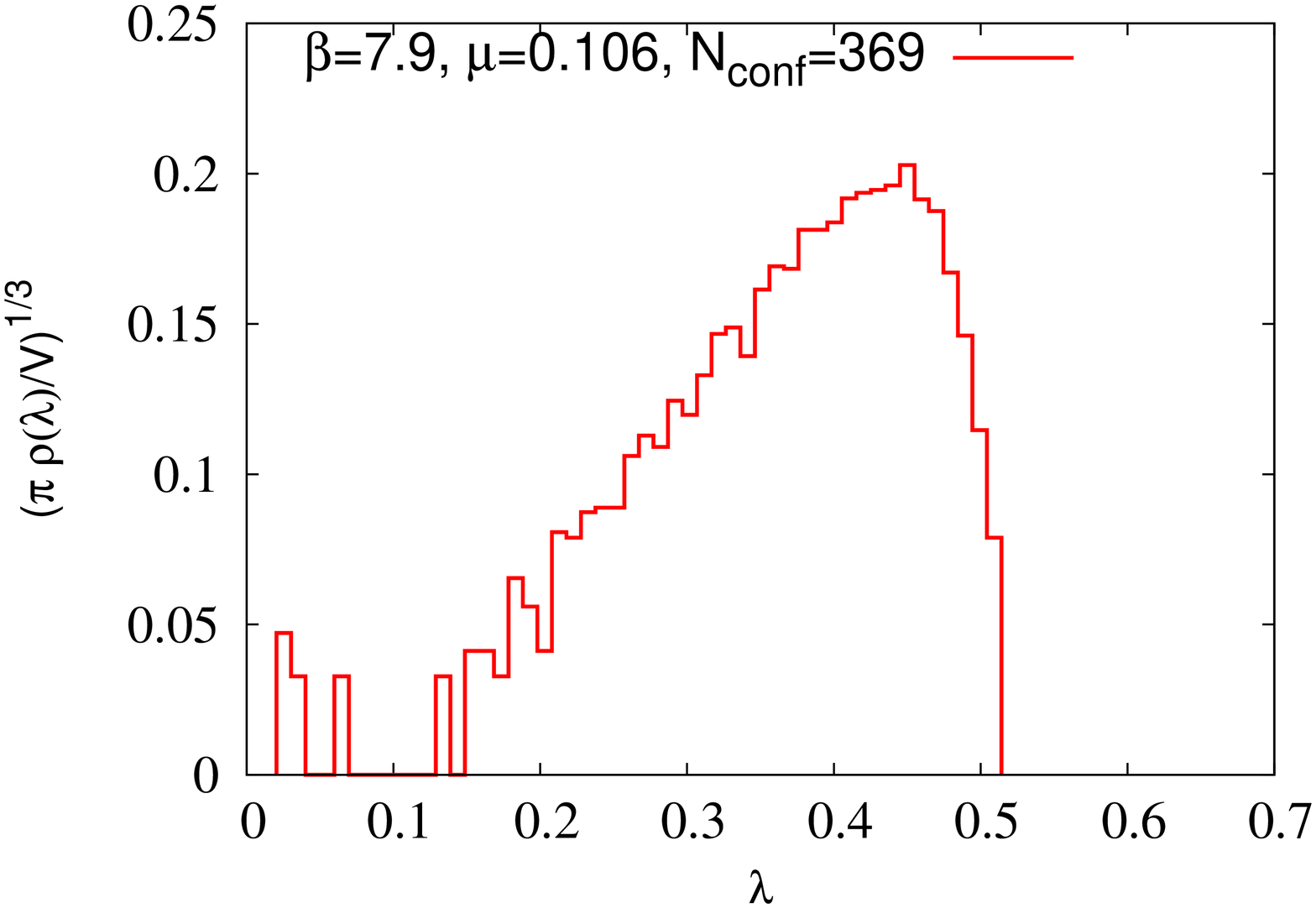}\includegraphics[angle=-90,width=9cm]{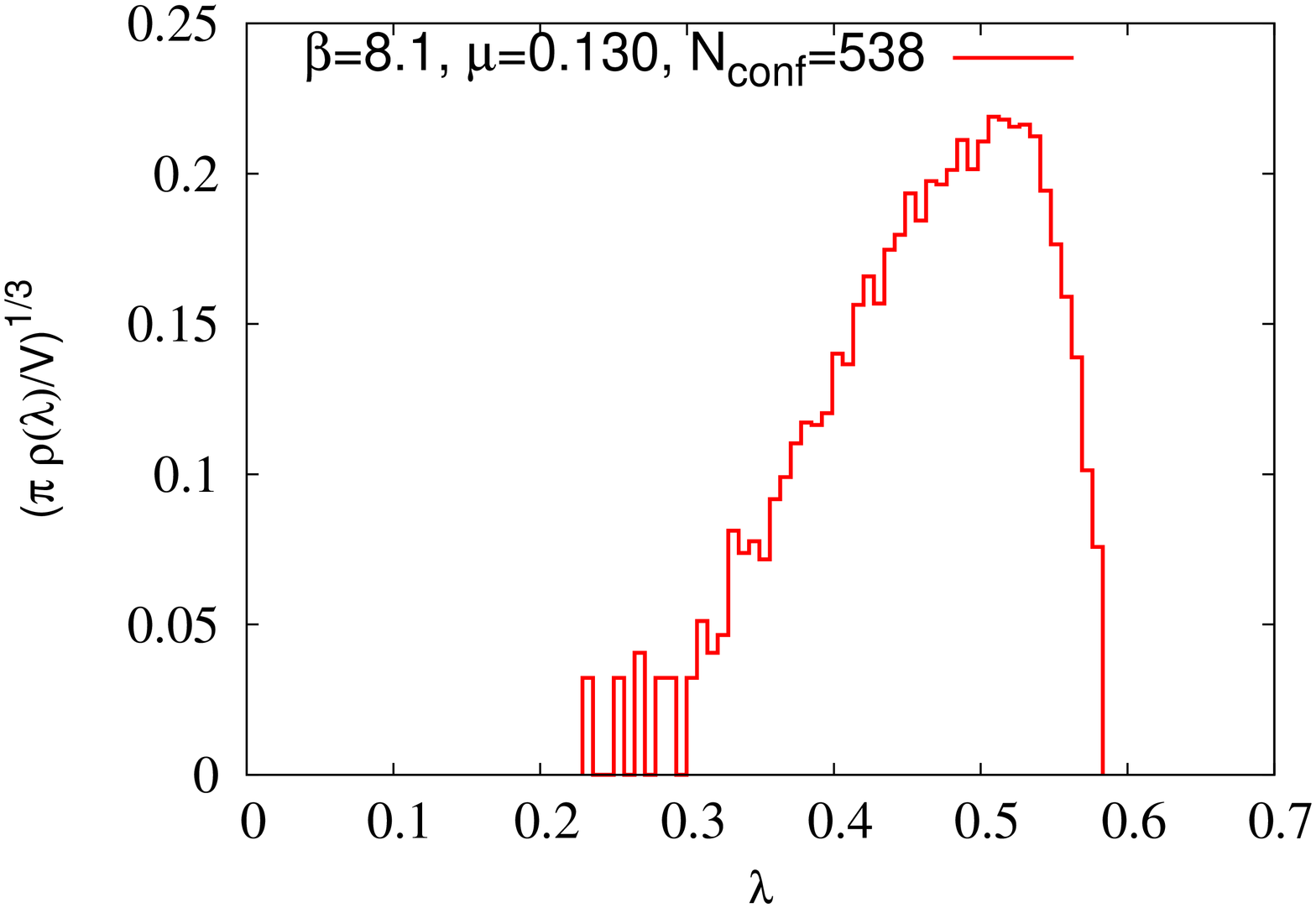}\\
\begin{flushleft}
\includegraphics[angle=-90,width=9cm]{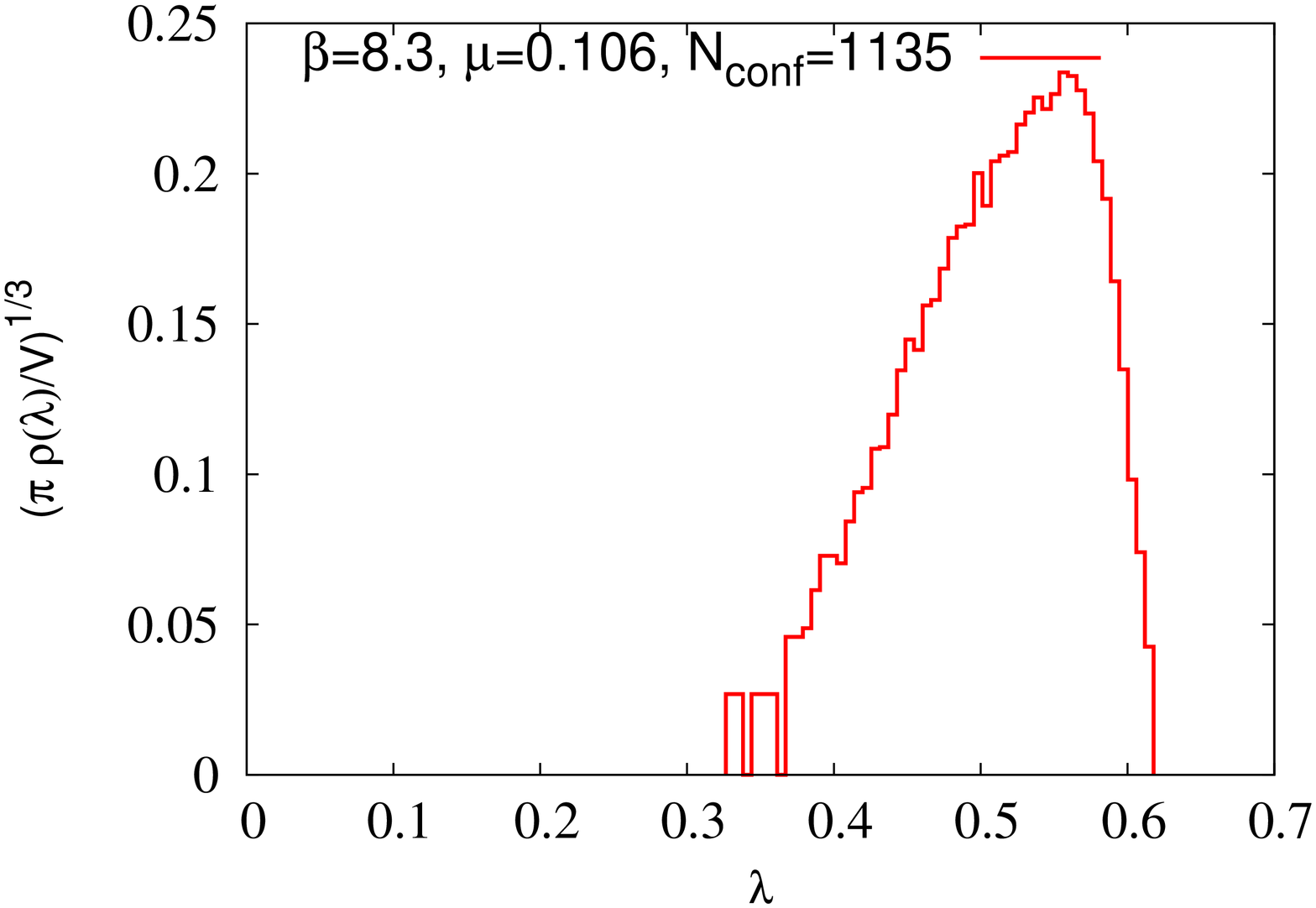}
\end{flushleft}
\caption{Histograms of the eigenvalues $\lambda$ of the overlap Dirac operator (\ref{massless_dirac}) in lattice units at different values of the inverse coupling constant $\beta$ which correspond to different temperatures in the range $220 \mev < T < 280 \mev$.}
\label{fig:hist_zero_field}
\end{figure*}

 The histograms of $\lambda$ at $\beta = 7.5$ ($T = 220 \mev$), $\beta = 7.7$, $\beta = 7.9$, $\beta = 8.1$ and $\beta = 8.3$ ($T = 280 \mev$) are shown on Fig.~\ref{fig:hist_zero_field}. In order to make the comparison with the chiral condensate easier, we rescale the eigenvalue density as in the Banks-Casher relation and plot the histograms in terms of $\lr{\pi \rho\lr{\lambda}/V}^{1/3}$. One can see that at $\beta = 7.5$ and $\beta = 7.7$ there is no gap in the spectrum of $\tilde{D}_0$, while at $\beta = 7.9$ the eigenvalue density moves away from zero and eventually a clearly visible gap forms. We interpret this gap opening as a signature of the restoration of chiral symmetry at the deconfinement temperature. From these data one can estimate that the transition temperature corresponds to some $\beta$ between $\beta = 7.9$ and $\beta = 8.1$.

 Finally, another argument which suggests that $T = 280 \mev$ corresponds to the deconfinement phase is that the topological charge fluctuates substantially at $T = 220\mev$ and at $T=280\mev$ it does not fluctuate at all. Also, independent lattice studies \cite{Bornyakov:2011,Bornyakov:2009} of the deconfinement phase transition with $N_f = 2$ flavors of improved Wilson fermions imply that at our value of $r_0 m_{\pi} = 1.24$ the deconfinement temperature is around $r_0 T \approx 0.5$, which is slightly lower than our value $r_0 T = 0.55$ for $T = 220 \mev$.

 We note that with our quark masses we most likely deal with a crossover rather than a finite-order phase transition \cite{Fodor:06:1, Fodor:06:2}. Therefore the temperatures of the deconfinement transition (which is typically determined from the peak of the Polyakov loop susceptibility) and of the chiral transition (which is extracted from the peak of the chiral susceptibility) might in general be different. Furthermore, the magnetic field might introduce some additional splitting between the two transitions or change their order \cite{Chernodub:10:2}. However, with our present statistics it is hardly possible to distinguish the two transitions or to determine their order. Therefore in the rest of the paper we simply refer to the range of temperatures in which the deconfinement and the restoration of chiral symmetry occur as the ``deconfinement transition'' for the sake of brevity.

 Given the two different values of quark charges $q_u=2 e/3$ and $q_d=-e/3$, the quantization of the magnetic field $B$ in our case is determined by the $d$ quark charge $q_d = -e/3=-|q|$,
\begin{eqnarray}
\label{mag_field_quantization}
 q B = 2 \pi  \frac{N_b}{L_s^2 a^2}  ,
\end{eqnarray}
where $L_s$ is the spatial lattice size and $N_b$ is an integer.

 Table \ref{tab:statistics} summarizes the number of gauge field configurations which we have used for different temperatures and magnetic fields.
The third and the fifth column give the total number of configurations which we have generated and the actual number of configurations which we have used to calculate the averages of the Polyakov loop and the chiral condensate. For the Molecular Dynamics (MD) part of the HMC algorithm, we used the Omelyan integrator \cite{Omelyan:03:1, Takaishi:05:1}. \
The length of MD trajectories ranged from $0.1$ to $0.2$ with the time step $0.001$ in the confinement phase. In the deconfinement phase, the MD trajectory length was close to one with a time step of $0.00305$. For all HMC processes, these parameters were tuned to reach the optimal acceptance rate of $0.8 \ldots 0.9$. In order to start HMC simulations, we have first performed several long runs (each consisting of several hundred HMC trajectories) at zero magnetic field using Zolotarev fermions \cite{Cundy:11:1}, since these will generate a similar ensemble as true overlap fermions but are significantly faster during the thermalization process. Then we continued thermalization with the overlap Dirac operator for around 100 additional HMC trajectories.
Ensembles at non-zero magnetic field were thermalized
for about 200 HMC trajectories starting from thermalized
configurations at zero magnetic field. Initial configurations for Monte-Carlo histories of our ``production runs'' shown in Fig.~\ref{fig:condensate_history}, Fig.~\ref{fig:ploop_history} and Fig.~\ref{fig:top_charge_history} correspond to the end of this thermalization process. In addition, during the $50$ subsequent HMC trajectories we have continued to tune the value of the improvement parameter in the gauge action. For this reason these $50$ first configurations have also been excluded from our analysis. It should be noted that due to the tuning of the improvement parameter at the beginning of each HMC process, these parameters slightly depend on the magnetic field. Since the magnetic field is an infrared parameter which does not affect renormalization, this dependence is a lattice artifact and should be negligible for sufficiently small lattice spacing. After completing the tuning of the action,  we started measuring each particular observable, and discarded the initial measurements if the average value had not yet stabilized. In total, our simulations took around $3 \cdot 10^6$ CPU-hours (mostly on $3 \, {\rm GHz}$ Intel Xeon CPUs). In order to estimate statistical errors and take into account autocorrelations in our data the blocked jackknife method was applied.

\begin{table}[h!]
\begin{center}
\begin{tabular}{|c|c|c|c|c|}
\hline
         &                      & No. of                &               & No. of   \\
   $N_b$ & $qB,\gevq$         & config.,              & $qB,\gevq$    &config.,\\
         &                      & $T=220\mev$           &               &$T=280\mev$\\
\hline
   0 & $0$     & $302/153$ & $0$     & $1333/1135$ \\
\hline
   1 & $0.044$ & $443/144$ & $0.068$ & $750/702$ \\
\hline
   5 & $0.22$  & $378/309$ & $0.34$  & $762/664$ \\
\hline
  10 & $0.44$  & $1396/1130$ & $0.68$  & $778/727$ \\
\hline
\end{tabular}
\caption{Summary of statistics. In the third and fifth column the first number is the total number of HMC trajectories in HMC processes and the second number is the actual number of configurations which were used to calculate the averages of the Polyakov loop and chiral condensate.}
\label{tab:statistics}
\end{center}
\end{table}

\section{Numerical results}
\label{sec:results}

\subsection{Chiral condensate and the distribution of Dirac eigenvalues}
\label{subsec:chiral_condensate}

\begin{figure}[h!tpb]
\includegraphics[angle=-90,width=8.5cm]{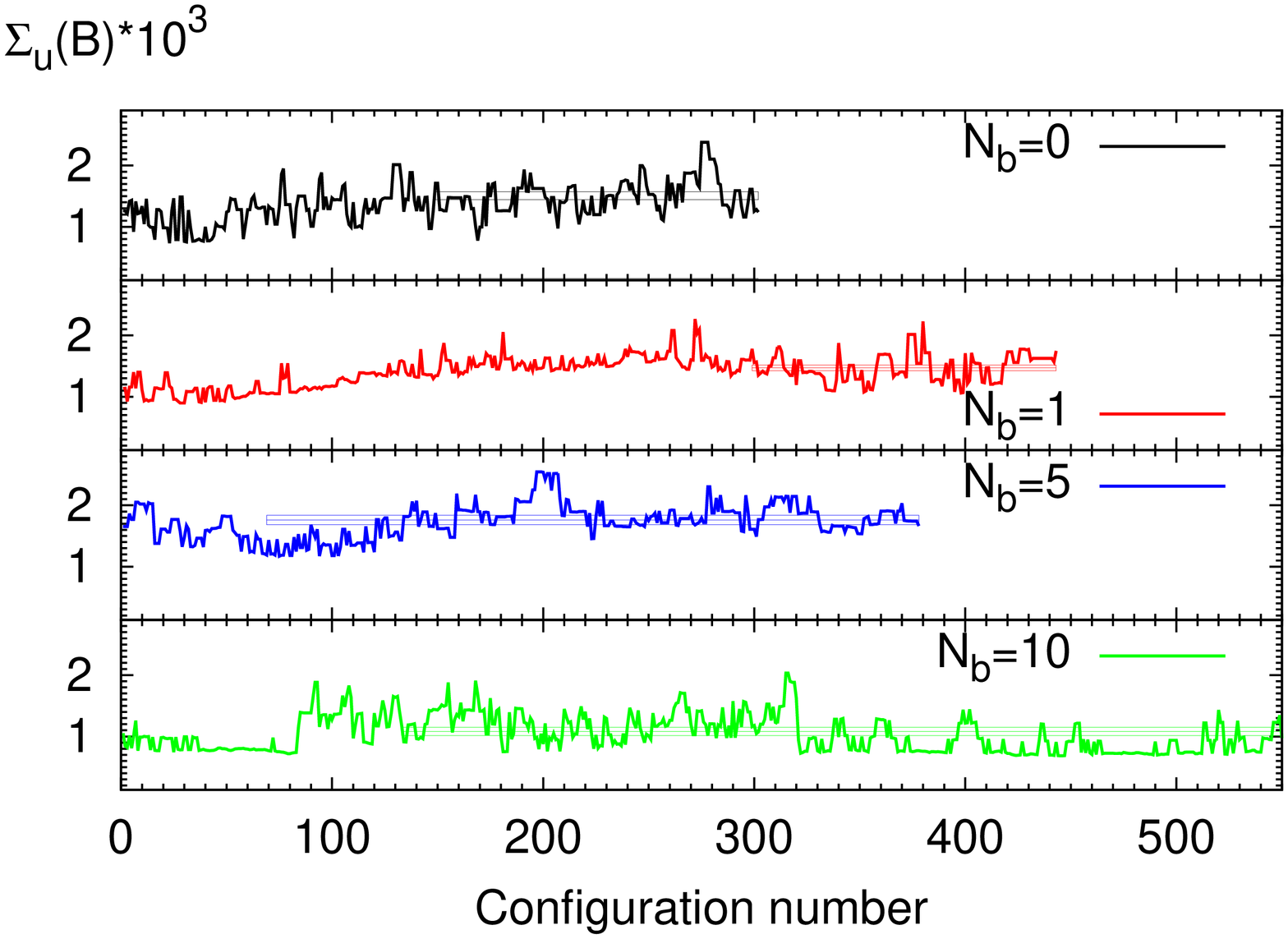}
\includegraphics[angle=-90,width=8.5cm]{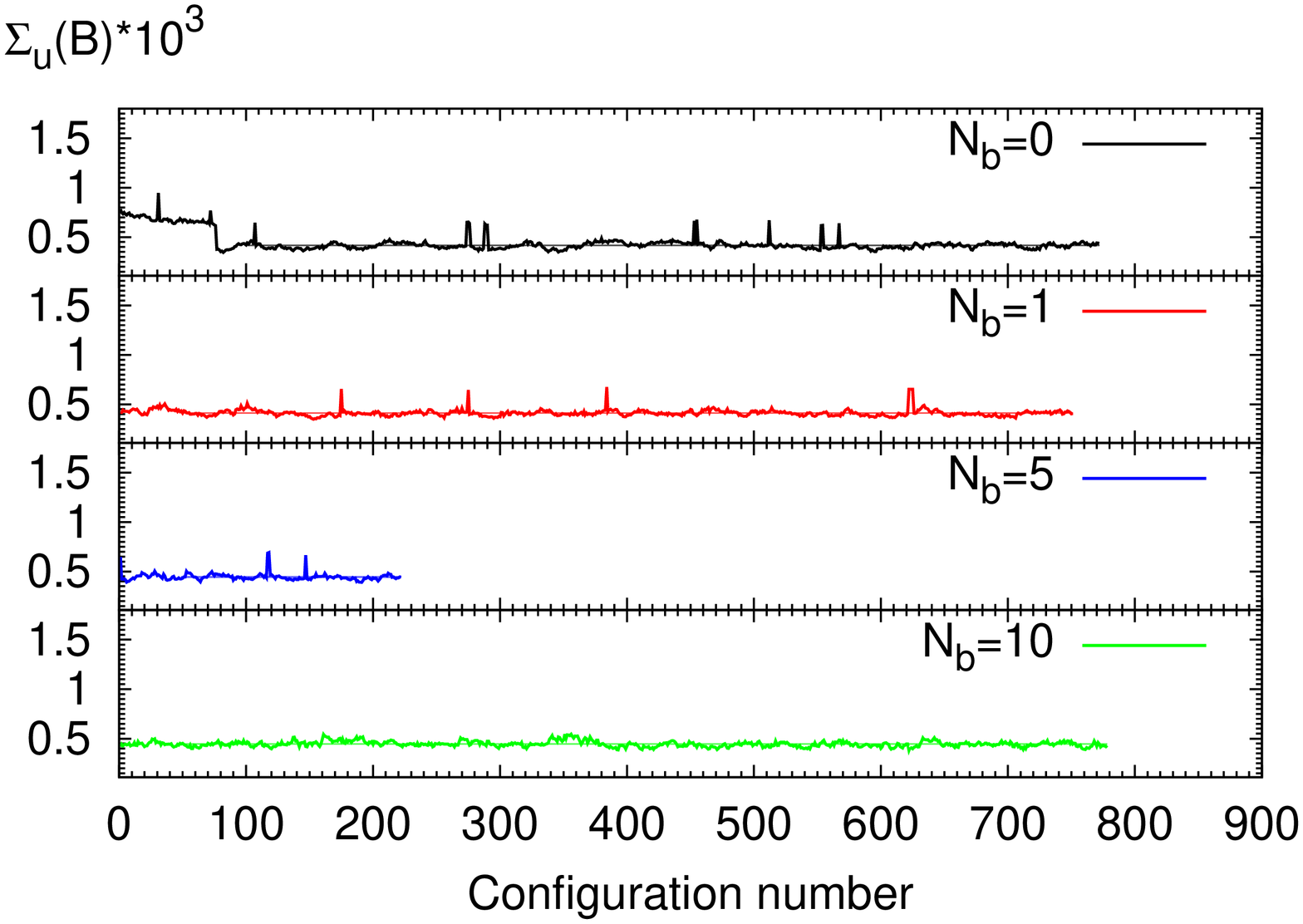}
\caption{Monte-Carlo histories of the chiral condensate (in lattice units) close to $T_c$  $(T=220\mev)$ (top) and in the deconfinement $(T=280\mev)$(bottom) regime for different magnetic fields. The length of the rectangles on the plot denotes the configurations which were used for statistical averaging, their central lines denote the corresponding expectation values and their height illustrates the statistical $1 \sigma$ error of these expectation values.}
\label{fig:condensate_history}
\end{figure}

 In the limit of zero quark mass the chiral condensate is an exact order parameter for spontaneous chiral symmetry breaking. In nature quarks are massive and QCD with physical quark masses as well as $N_f=2$ QCD with light enough quarks (as in our case) exhibit a crossover instead of a phase transition at finite temperature \cite{Fodor:06:1, Fodor:06:2}. Therefore, the chiral condensate can serve only as an approximate order parameter, which still decreases near the deconfinement temperature, although not to zero. In this work we study the $u$-quark condensate
\begin{eqnarray}
\label{condensate_def}
 \Sigma_u = -\rho^{-1} \vev{\bar{\psi}_u \lr{1 - D_0/2} \psi_u } ,
\end{eqnarray}
where the factor $\lr{1 - D_0/2}$ is a finite-spacing lattice correction \cite{Chandrasekharan:99:1} and $\rho$, the negative Wilson mass parameter in the definition of overlap operator, is the usual wave function renormalization factor for overlap fermions. Since the $u$-quark has the largest electric charge, this condensate is most sensitive to the magnetic field. In particular, one can expect that if inverse magnetic catalysis is observed for $\Sigma_u$, it will be also observed for the condensate $\Sigma_d$ of $d$-quarks as well as for the total condensate $\Sigma = \Sigma_u + \Sigma_d$. Indeed, since for a fixed gauge field configuration the magnetic field tends to increase the density of low-lying Dirac eigenvalues \cite{Endrodi:13:new}, inverse magnetic catalysis is an interplay of two competing effects. On the other hand, the increase of the density of low-lying Dirac eigenvalues increases the condensate (which is related to the density of Dirac eigenvalues by the Banks-Casher relation \cite{Banks:80:1}). On the other hand, a larger density of low-lying eigenvalues decreases the fermionic determinant and thus decreases the relative weight of this configuration in an equilibrium statistical ensemble. Inverse magnetic catalysis is observed if the effect of decrease of statistical weight due to the magnetic field is larger than the corresponding (valence) increase of the condensate \cite{Endrodi:13:new}. The first effect is most prominent if we consider the condensate of $u$-quarks, which have the maximal charge $q_u = 2 e/3$. The effect of the decrease of the fermionic determinant is independent of whether we calculate the condensate of the $u$- or $d$-quarks. If the decrease of the fermionic determinant wins over the increase of the condensate already for the quark with the maximal charge, for smaller quark charges this effect will be even stronger. Therefore if inverse magnetic catalysis is observed for the condensate of the $u$-quark, for the condensate of the $d$-quark or for the sum of both condensates it will be even more pronounced.

\begin{figure}[h!tpb]
\center{\includegraphics[angle=-90,width=9cm]{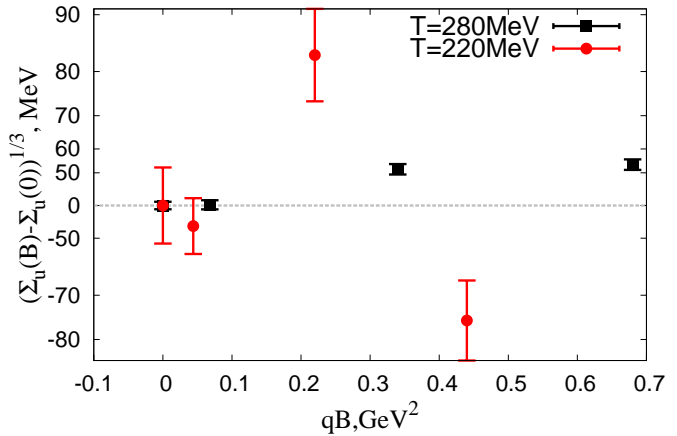}}
\caption{The difference $\Sigma\lr{B,T} - \Sigma\lr{0,T}$ in the values of the chiral condensate at zero and nonzero external magnetic field for $T=220\mev$ and $T=280\mev$. In order to facilitate reading the plot in physical units, the scale of the vertical axis is chosen to be the cubic scale.}
\label{fig:condensate_average}
\end{figure}

 Monte-Carlo histories of the $u$-quark condensate (\ref{condensate_def}) at $T=220\mev$  and $T=280\mev$ and for different magnetic field strengths are shown in Fig.~\ref{fig:condensate_history} (we give all results in lattice units). The condensate was calculated using $64$ Gaussian stochastic estimators. The length of the rectangles on the plot mark the part of the full HMC history which was used for statistical averaging, their central line denotes the expectation values and their width is the corresponding statistical $1 \, \sigma$ error. We note that above the deconfinement temperature the chiral condensate thermalizes much faster than below. Autocorrelation times are about $10 - 20$ HMC trajectories for $T=280\mev$ and about $30$ HMC trajectories for $T = 220\mev$. One can also see from Fig.~\ref{fig:condensate_history} that during thermalization the chiral condensate exhibits some long-range fluctuations.

 In Fig.~\ref{fig:condensate_average} we show the dependence of the chiral condensate in physical units on the magnetic field. Since external magnetic field does not induce any additional divergences in the chiral condensate \cite{Bali:13:1, Bali:14:1}, in order to obtain the finite physical answer we subtract the value of the condensate at zero magnetic field and the same temperature from our results. Based on our data at $T = 220 \mev$ we see some indications that the chiral condensate first increases with magnetic field up to $qB=0.2 \gevq$, and clear evidence that it decreases at larger values of $qB$. This means that a sufficiently strong magnetic field tends to restore the chiral symmetry and the theory approaches the deconfinement regime in which the chiral symmetry is restored. Such behaviour thus favors the inverse magnetic catalysis scenario. At $T = 280 \mev$ the condensate slowly increases.

\begin{figure*}[pH]
 \includegraphics[angle=-90,width=8cm]{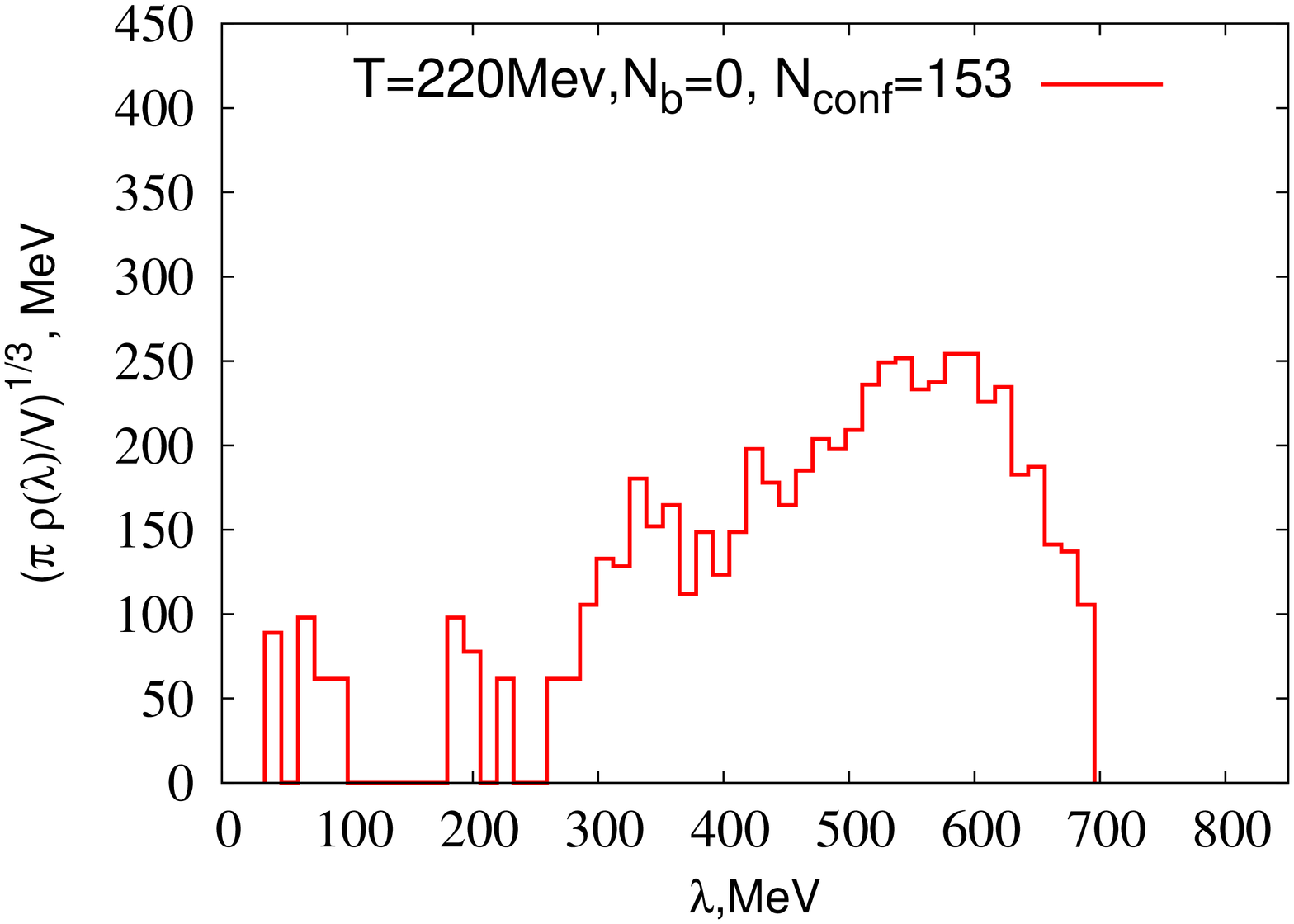}\includegraphics[angle=-90,width=8cm]{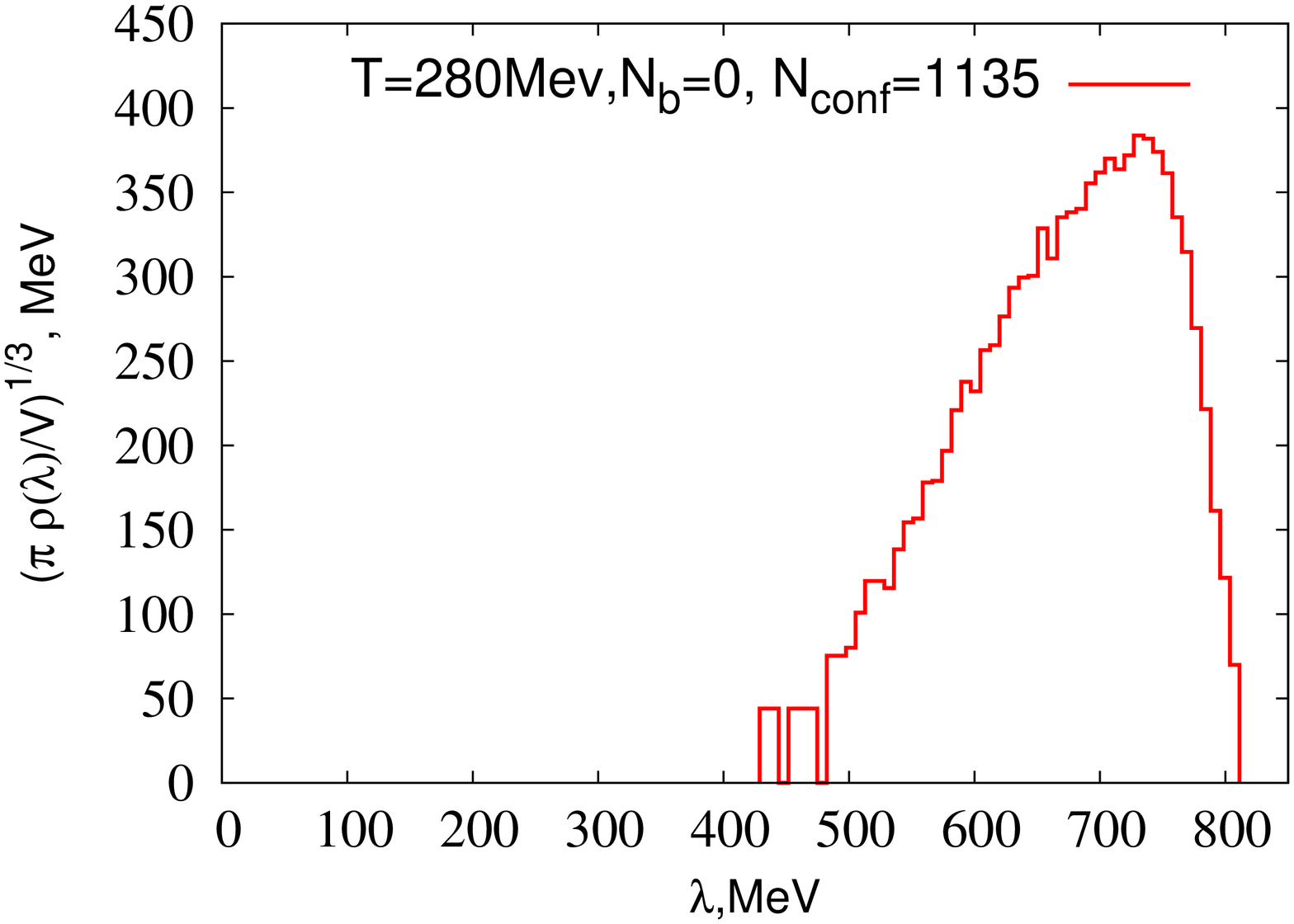}\\
 \includegraphics[angle=-90,width=8cm]{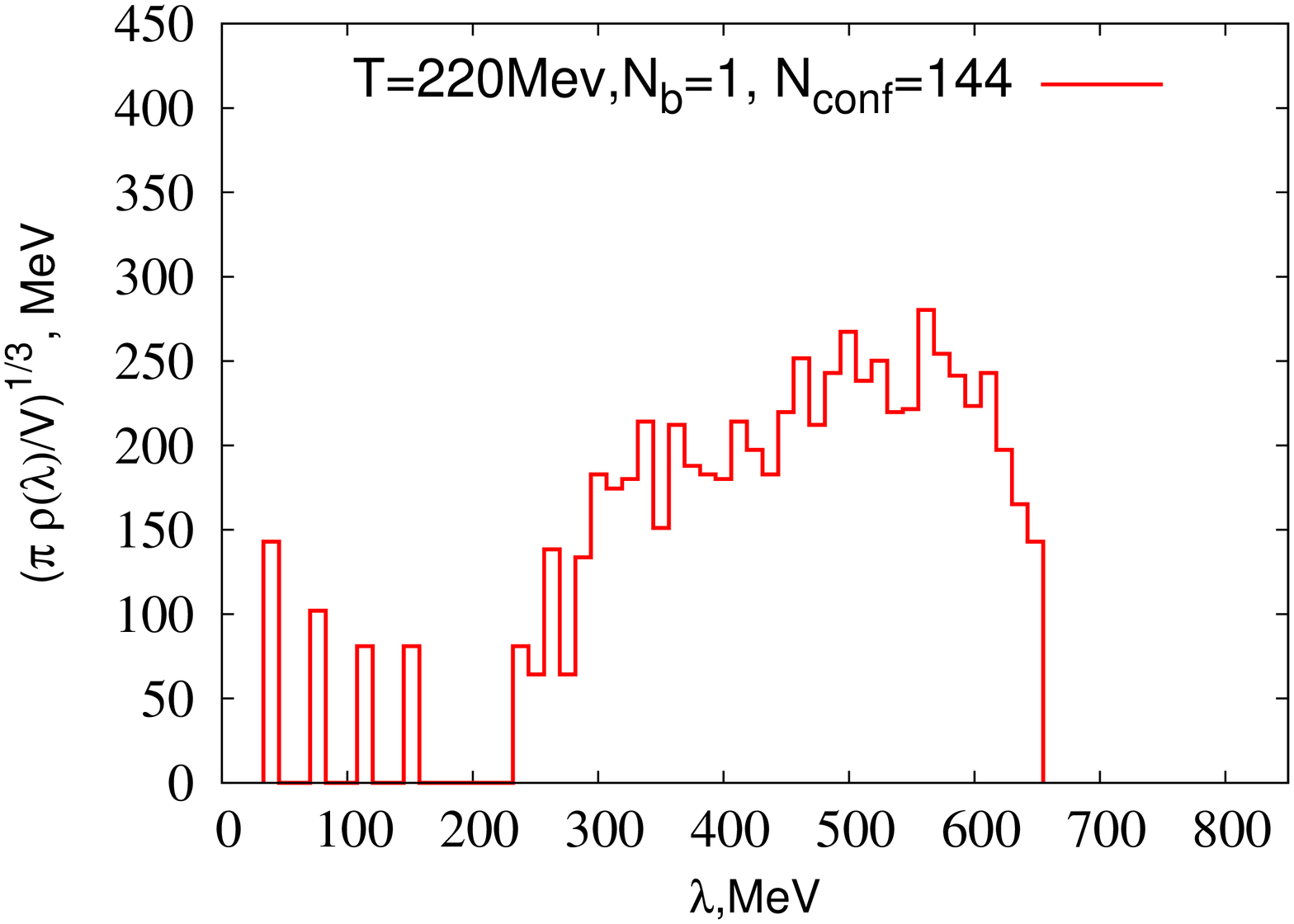}\includegraphics[angle=-90,width=8cm]{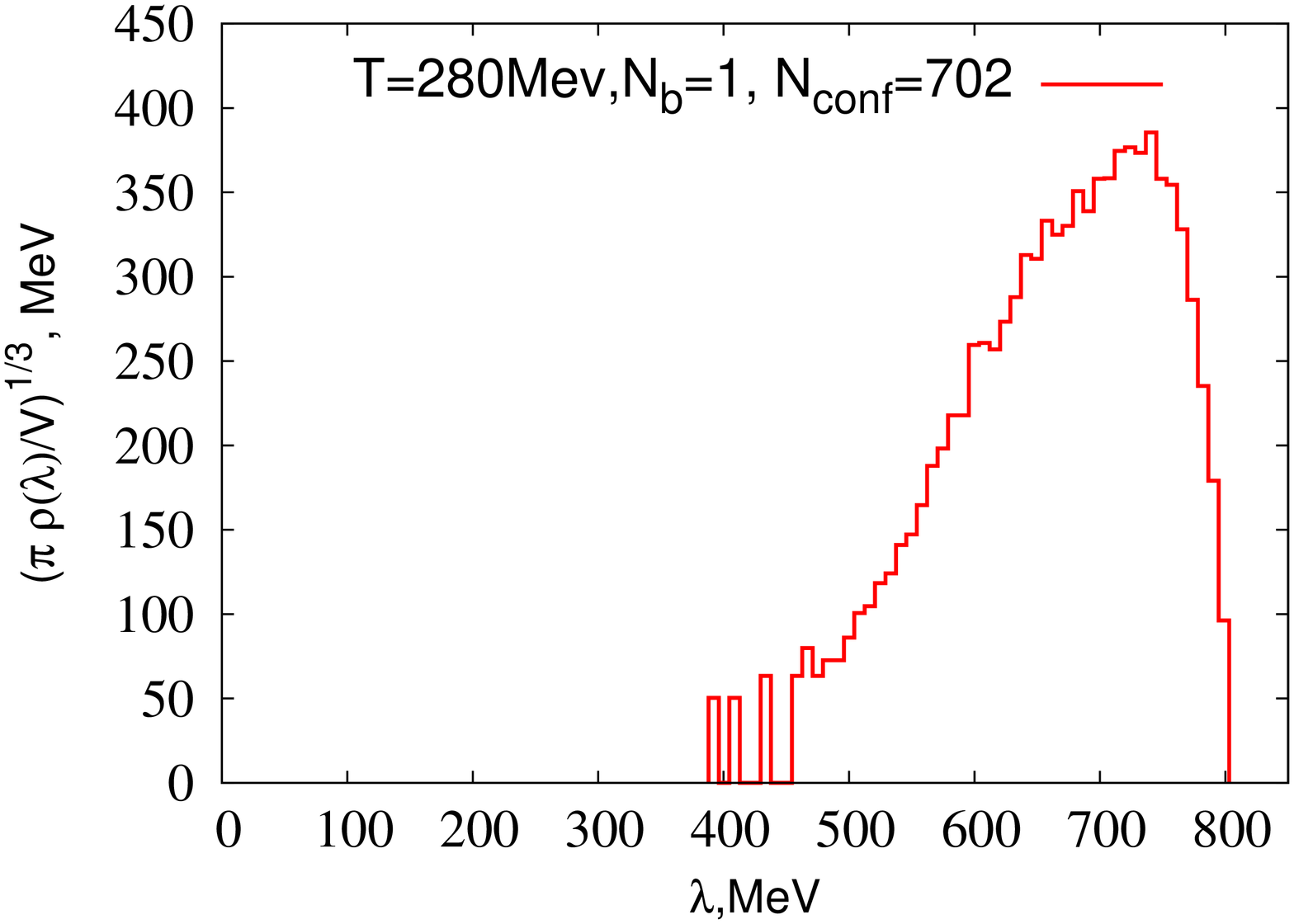}\\
 \includegraphics[angle=-90,width=8cm]{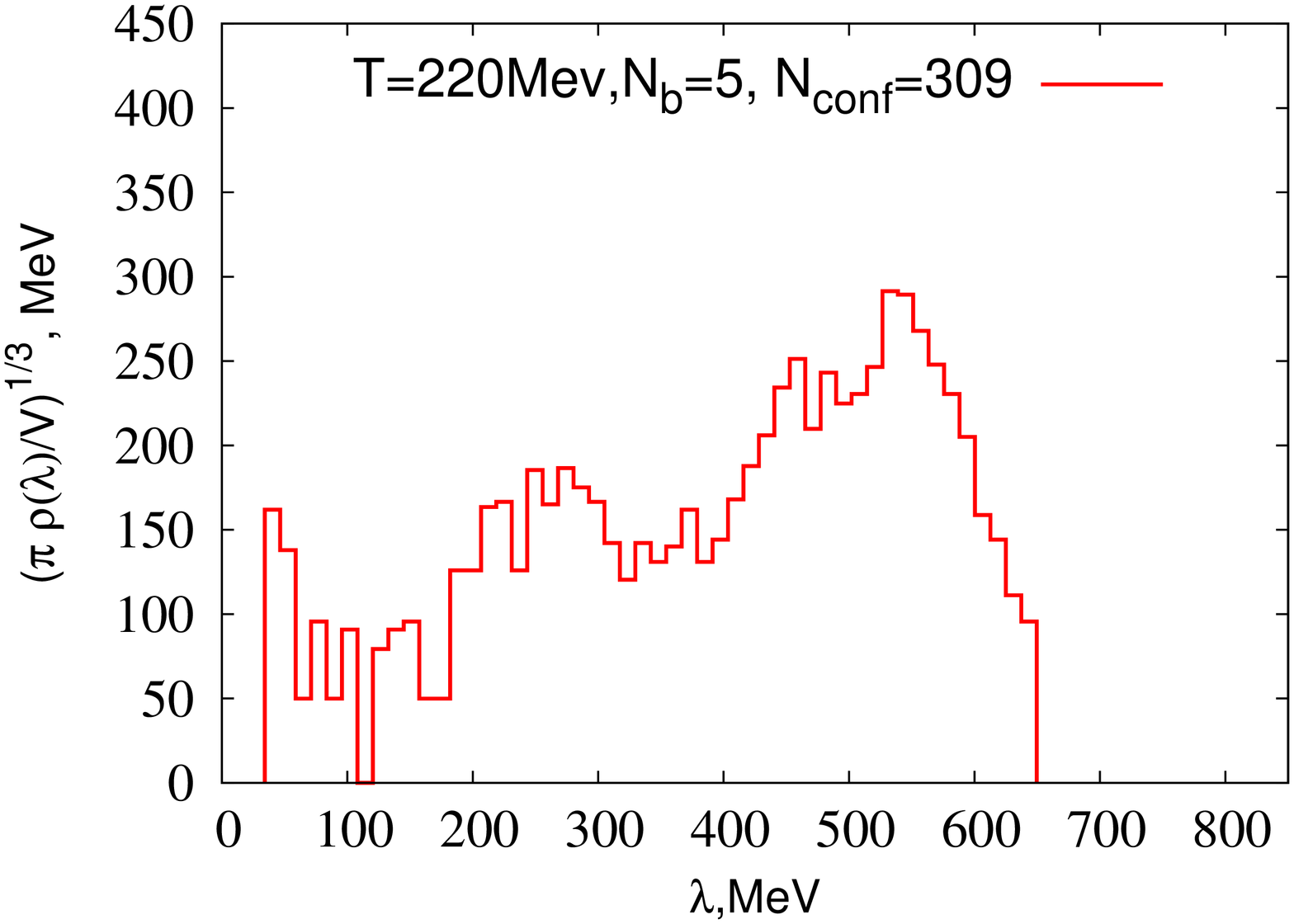}\includegraphics[angle=-90,width=8cm]{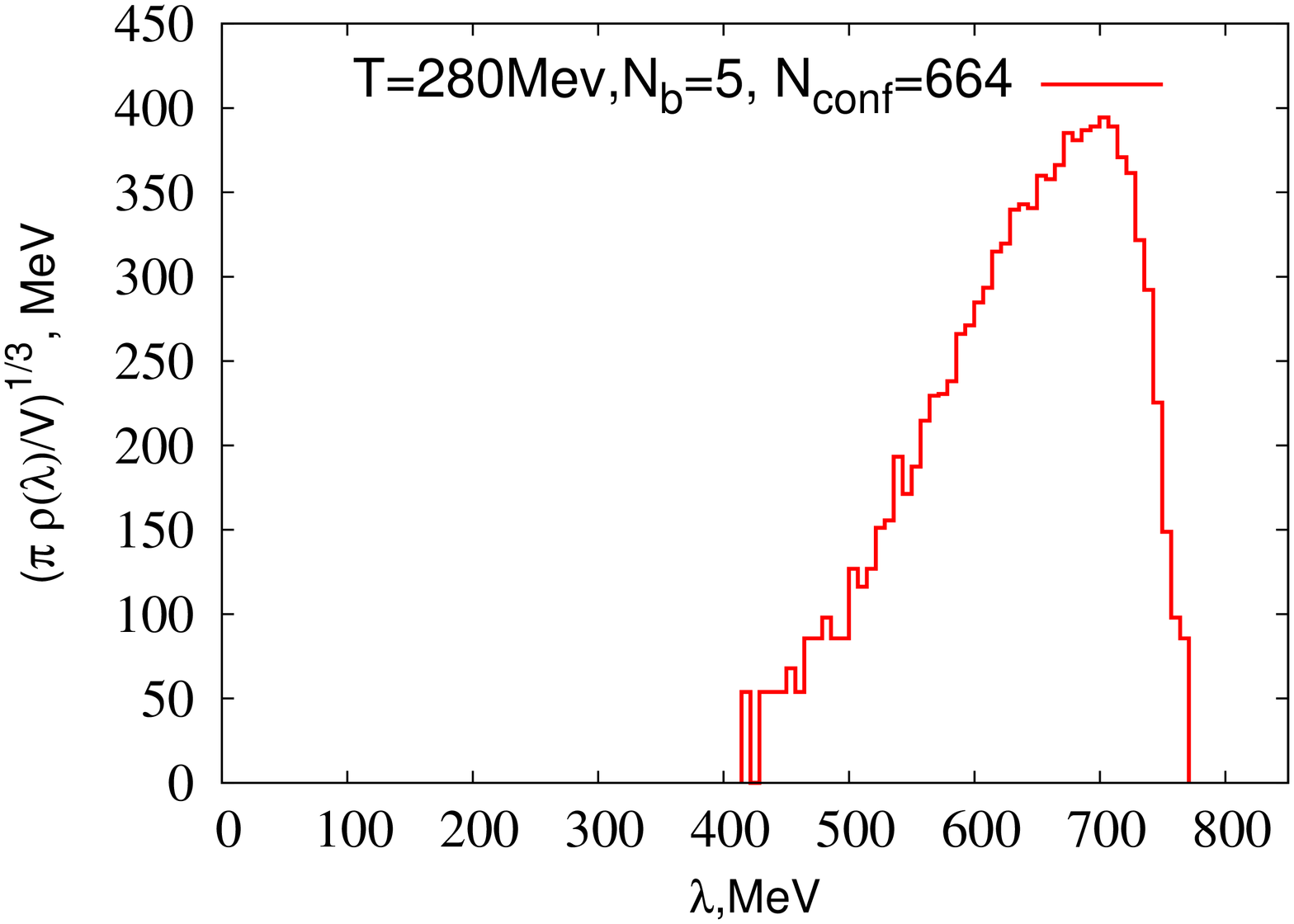}\\
 \includegraphics[angle=-90,width=8cm]{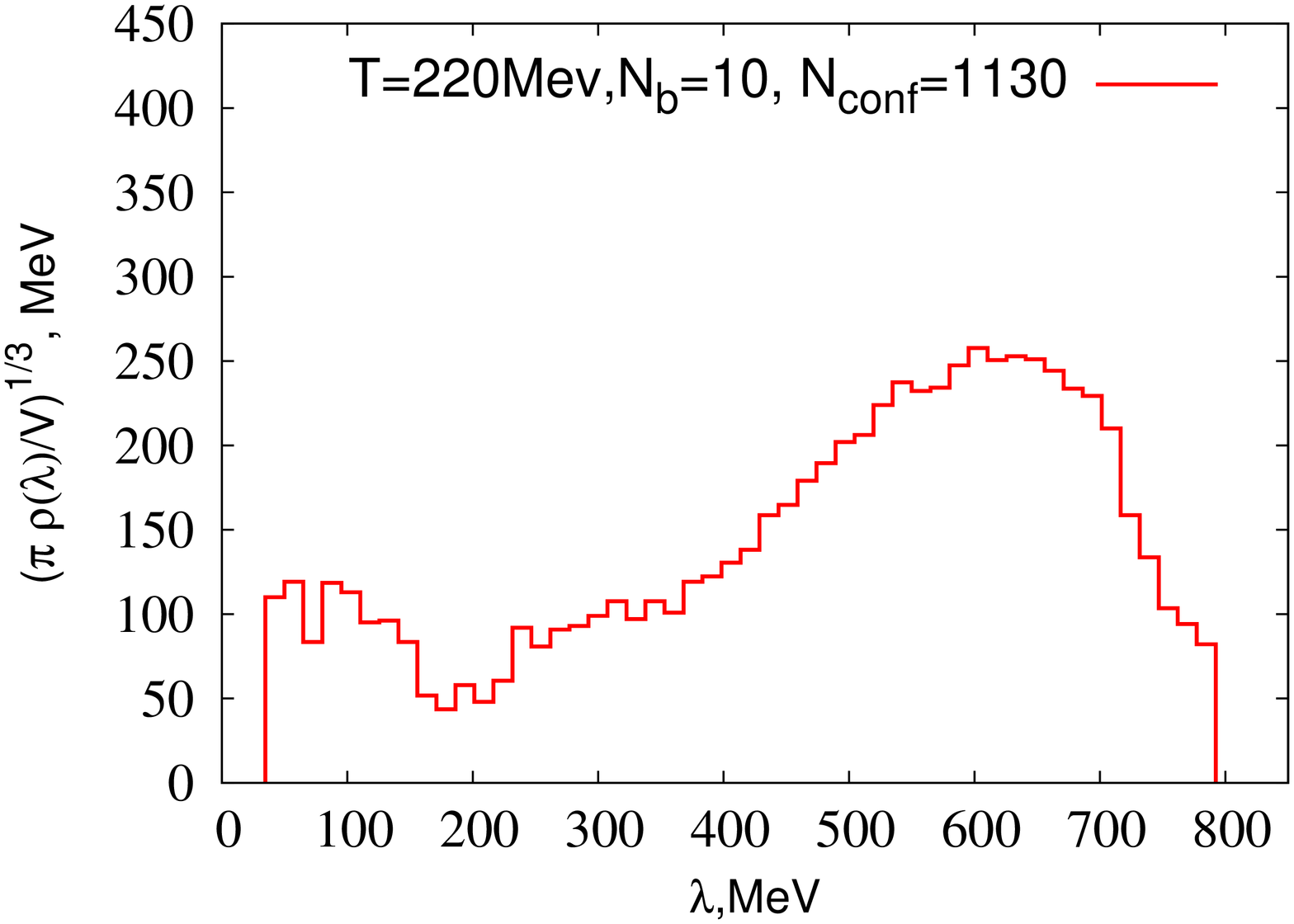}\includegraphics[angle=-90,width=8cm]{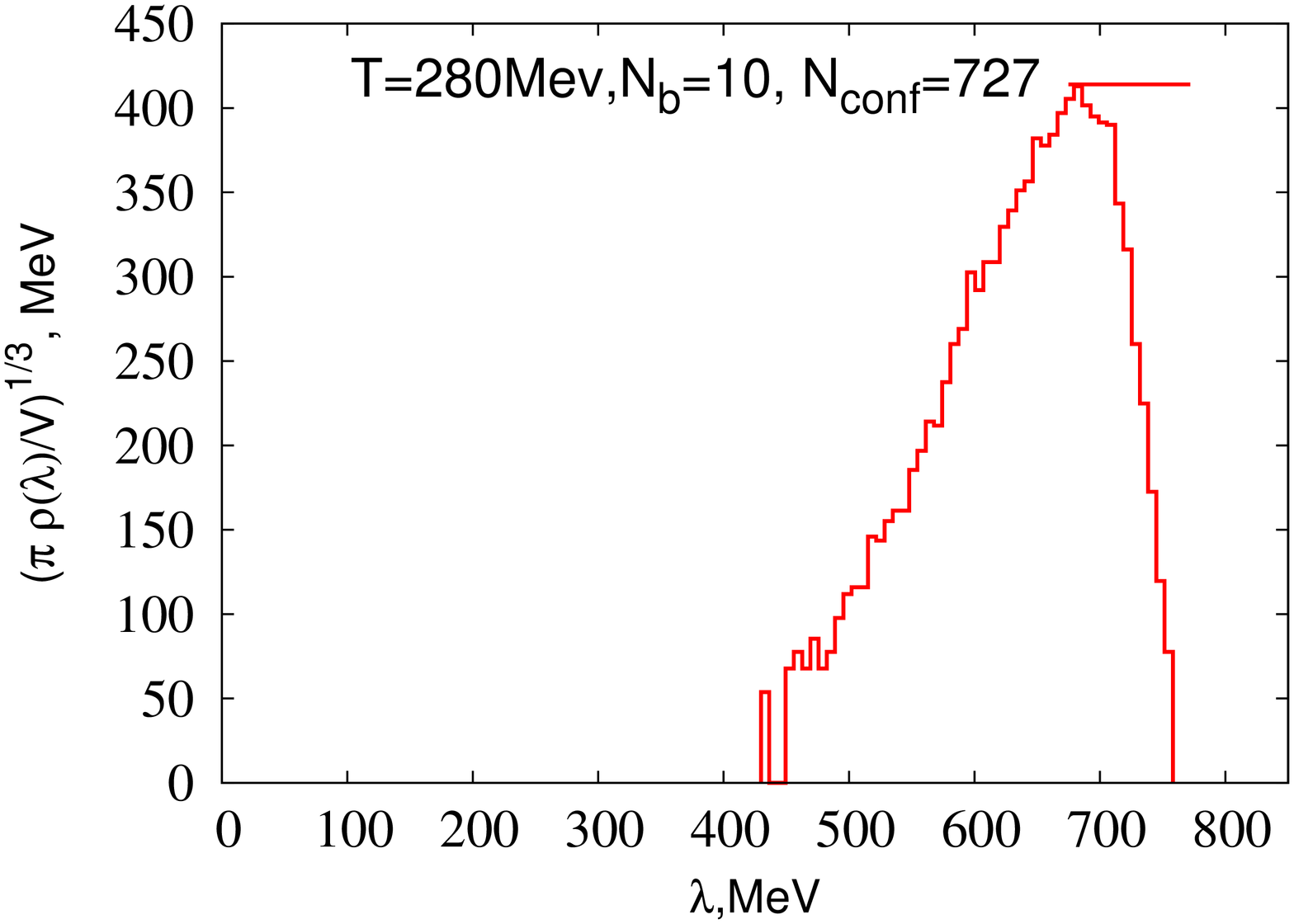}\\
\caption{Histograms of the eigenvalues $\lambda$ of the operator (\ref{massless_dirac}) at temperatures $T = 220 \mev$ ($\beta = 7.5$, on the left) and $T = 280 \mev$ ($\beta = 8.3$, on the right) and at different values of the magnetic field flux.}
 \label{fig:hist_mag_field}
\end{figure*}

 To obtain a more complete picture of the restoration of chiral symmetry, we have also studied how the spectrum of low-lying eigenvalues $\lambda$ of the operator (\ref{massless_dirac}) for the $u$-quark changes with magnetic field at $T = 220 \mev$ and at $T = 280 \mev$. The corresponding histograms of $\lambda$ are shown on Fig.~\ref{fig:hist_mag_field}. Again, we plot the eigenvalue density rescaled as $\lr{\pi \rho\lr{\lambda}/V}^{1/3}$. One can see that for $T = 220 \mev$ ($\beta = 7.5$) the eigenvalue density becomes somewhat larger at small $\lambda$ for $N_b = 1$ and $N_b = 5$, but for $N_b = 10$ it decreases significantly, which again indicates that chiral symmetry tends to be restored at sufficiently high magnetic fields. At $T = 280 \mev$ the spectrum of $\lambda$ does not change significantly with magnetic field. This behaviour also completely agrees with the dependence of the chiral condensate (\ref{condensate_def}) on the magnetic field (see Fig.~\ref{fig:condensate_average}).

 Recent studies of the Regensburg group \cite{Endrodi:12:prd} revealed that in the vicinity of the deconfinement transition the chiral condensate first increases for not very high values of the magnetic field, and then starts decreasing. Taking into account that in our case the ensemble at $\beta=7.5$ ($T=220\mev$) is close to deconfinement transition, our results are in agreement with \cite{Endrodi:12:prd} except for the fact that our masses are much larger than in \cite{Endrodi:12:prd}. For the same value of the pion mass which we have used in our simulation, the simulations of \cite{Endrodi:12:prd} have actually reproduced magnetic catalysis rather than inverse magnetic catalysis. This suggests that the use of chiral lattice fermions strengthens the signatures of inverse magnetic catalysis.

\subsection{Polyakov loop}
\label{subsec:polyakov_loop}

 The Polyakov loop serves as an order parameter for QCD in the limit of infinitely heavy quarks. Similarly to the chiral condensate, it still can be considered as an approximate order parameter for the crossover transition in QCD with finite quark masses which increases from confinement to deconfinement.

\begin{figure}[h!tpb]
\includegraphics[angle=-90,width=8.5cm]{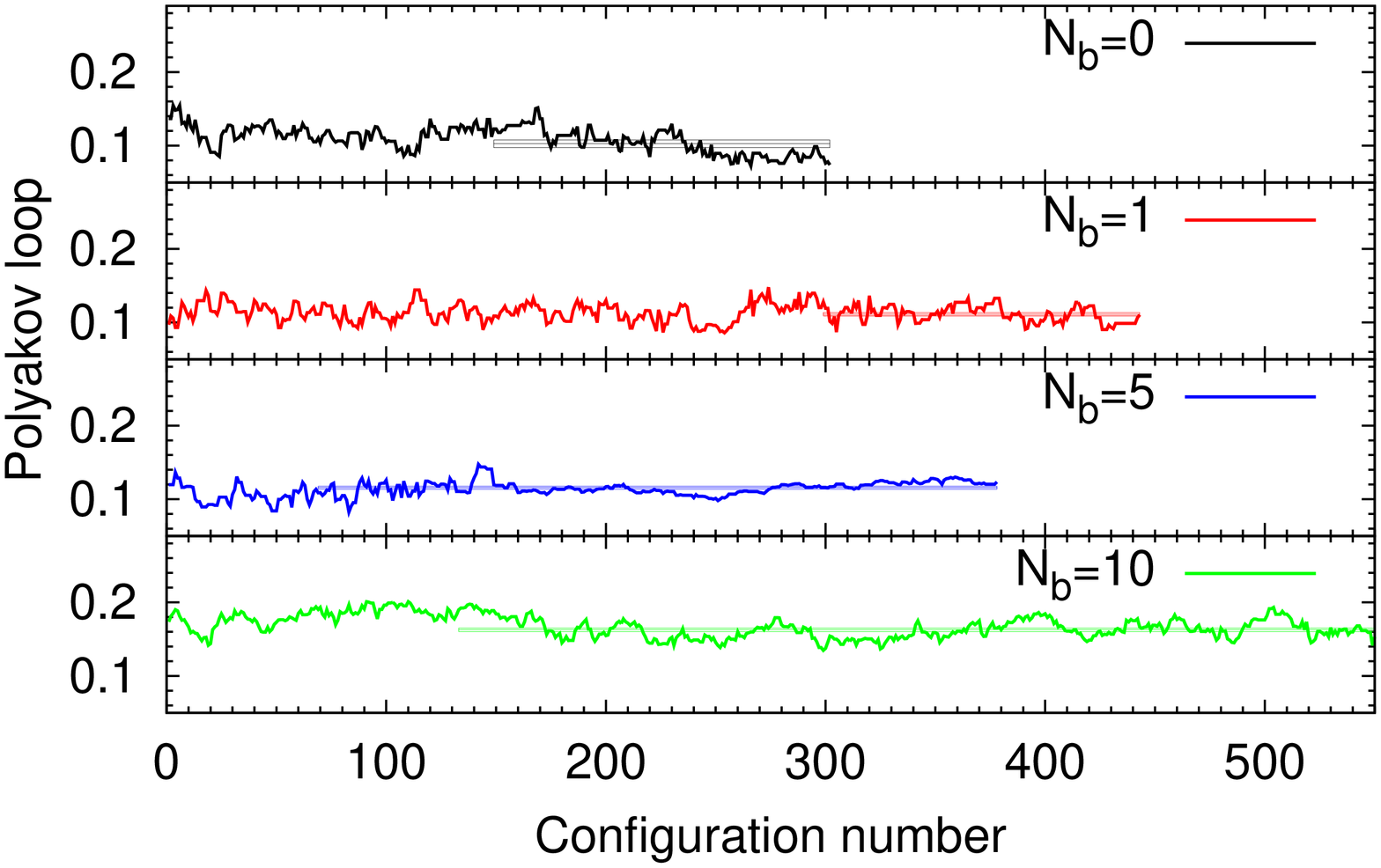}
\includegraphics[angle=-90,width=8.5cm]{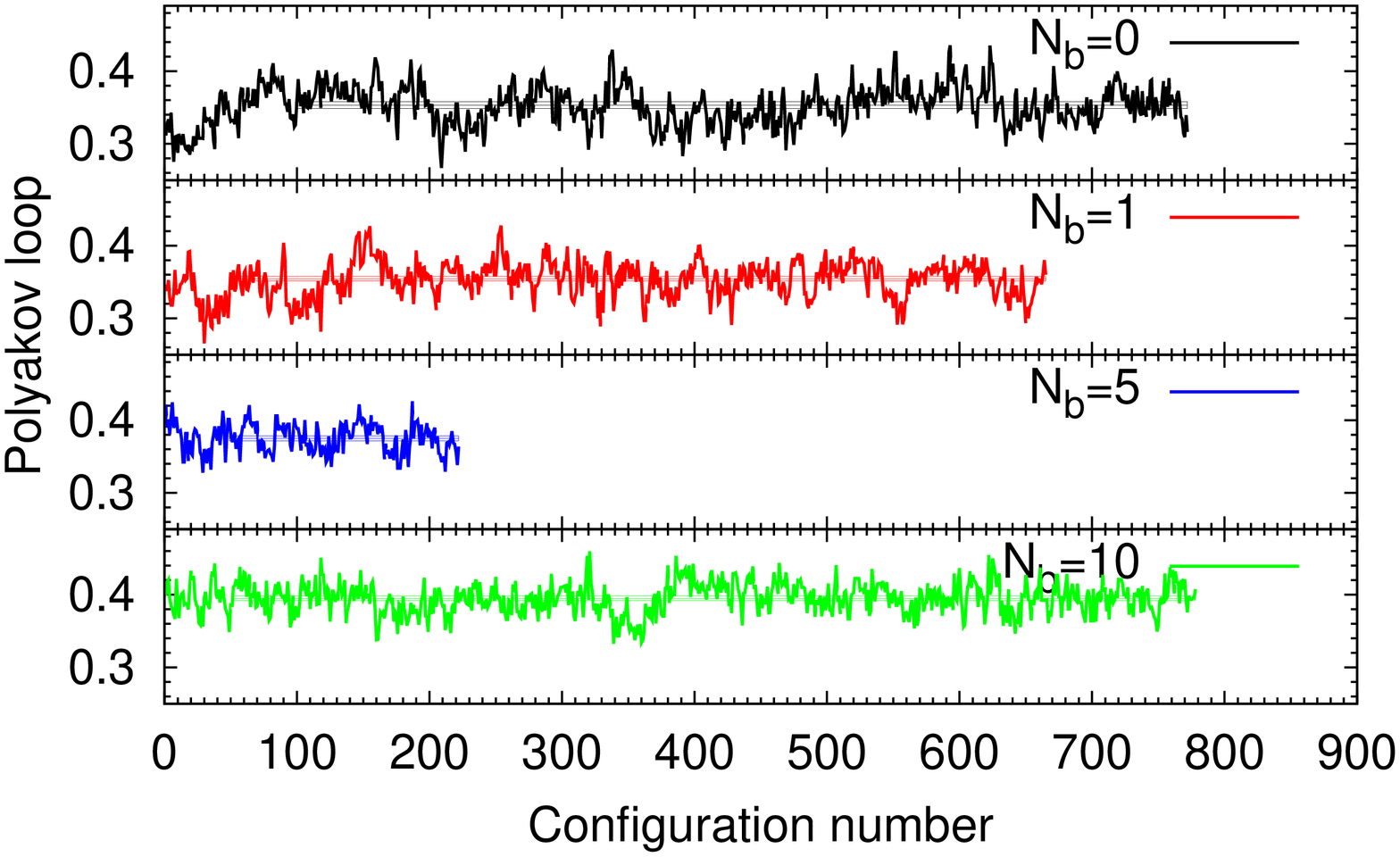}
\caption{Monte-Carlo histories of the Polyakov loop for $T=220\mev$ and $T=280\mev$ and different magnetic fields. The length of the rectangles on the plot denotes the configurations which were used for statistical averaging, their central lines denote the corresponding expectation values and their height illustrates the statistical $1 \sigma$ error of these expectation values.}
\label{fig:ploop_history}
\end{figure}

 Monte-Carlo histories of the Polyakov loop at $T=220\mev$ and $T=280\mev$ are shown in Fig.~\ref{fig:ploop_history} for different magnetic fields. Again, the length of the rectangles denotes the data set which was used for statistical averaging, their central line denotes expectation values and their width corresponds to the statistical error of the expectation value. The autocorrelation time for the Polyakov loop is around $20$ HMC trajectories for $T=280\mev$ and from $10$ to $30$ HMC trajectories for $T = 220 \mev$.

\begin{figure}[h!tpb]
\center{\includegraphics[angle=-90,width=9cm]{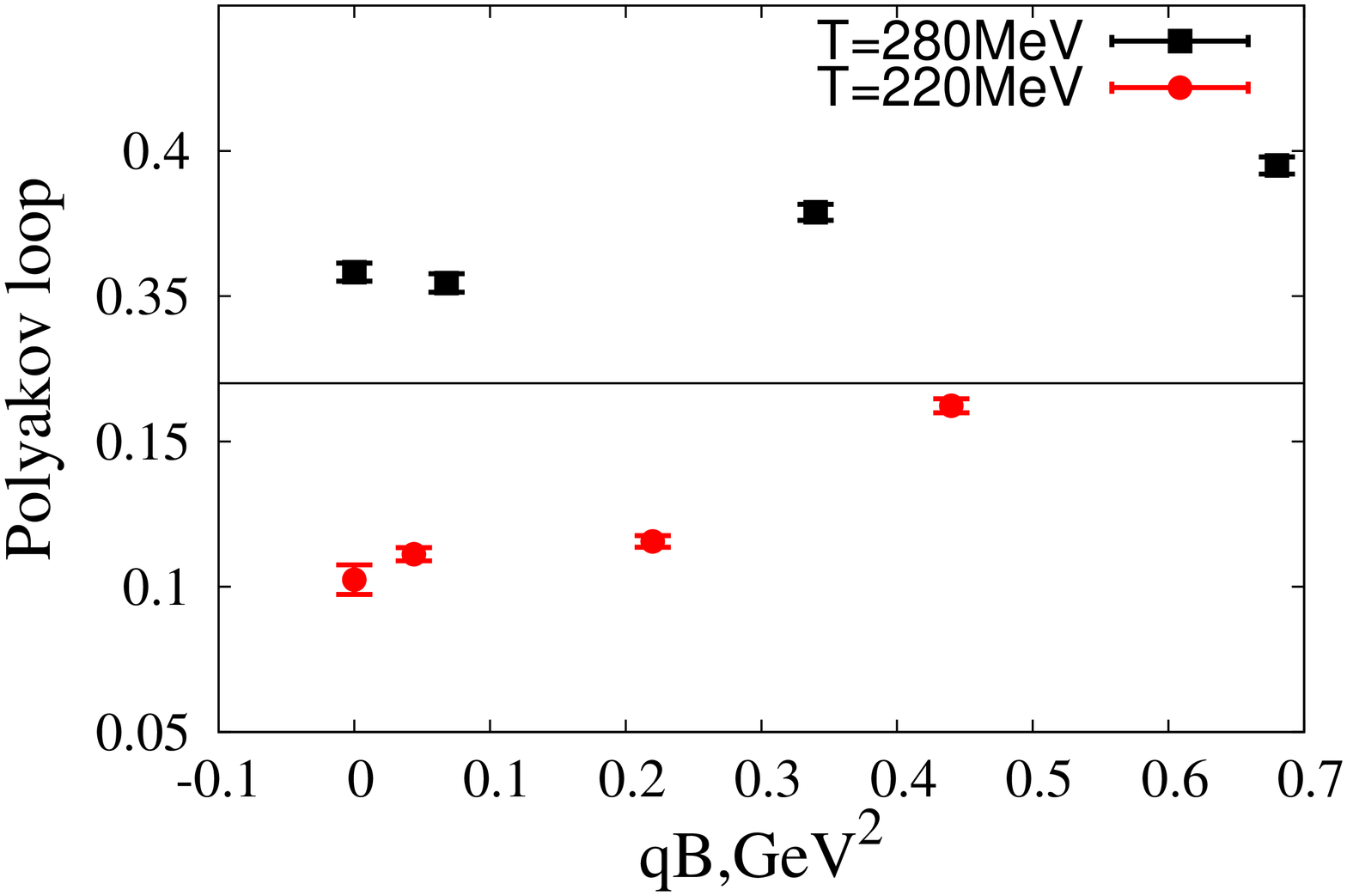}}
\caption{The Polyakov loop as a function of magnetic field for $T=220\mev$ and $T=280\mev$.}
\label{fig:ploop_average}
\end{figure}

 The dependence of the expectation values of the Polyakov loop on the magnetic field is shown in Fig.~\ref{fig:ploop_average}. One can see that the Polyakov loop grows with magnetic field for sufficiently large field strength ($q B \gtrsim 0.3 \gevq$). This means that the quark free energy decreases and the theory approaches the deconfinement regime, which again indicates that we observe inverse magnetic catalysis. Smaller magnetic fields do not change the Polyakov
loop within the statistical error range.  Of course, much more precise statements on the shift of the deconfinement temperature can be made if one considers the Polyakov loop susceptibility (see e.g. \cite{D'Elia:10}), however, our statistics is far too limited to perform such an analysis.

 Similar results for the Polyakov loop were also obtained in the papers \cite{Endrodi:12:jhep, Muller-Preussker:13}, where the Polyakov loop was observed to increase with magnetic field at all temperatures, and in \cite{D'Elia:10}, where the magnetic field slightly increased the Polyakov loop in the deconfinement regime and decreased it in the confinement regime.

\subsection{Topological charge}
\label{subsec:top_charge}

 Many interesting properties of the quark-gluon plasma in an external magnetic field such as anomalous transport phenomena \cite{Landsteiner:12:1, Basar:12:1, Kharzeev:08:2} and local $\mathcal{CP}$ violation \cite{Kharzeev:08, Chao:13, Yu:14:1} are believed to be related to topological transitions. Thus it is important to understand how the magnetic field changes the topological content of the QCD vacuum. It should be stressed that a magnetic field can affect topology only due to the back-reaction of dynamical quarks on the gauge fields, which can only be observed in lattice simulations with dynamical chiral fermions.

 For this reason we have also studied the influence of the magnetic field on the fluctuations of topological charge. On the lattice the index theorem allows us to calculate the topological charge by counting the  number of zero modes of the overlap Dirac operator. This is a natural way to explore topology of the gauge fields on the lattice because it does not require any smearing procedure and further approximation to obtain integer numbers.

\begin{figure}[h!tpb]
\center{\includegraphics[angle=-90,width=9cm]{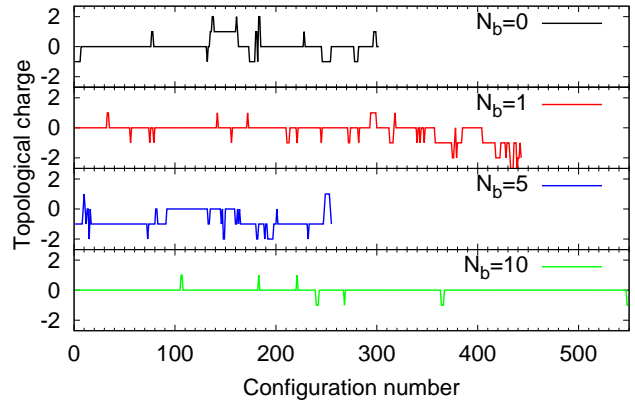}}
\caption{Monte-Carlo histories of the topological charge in the confinement regime ($T = 220 \mev$).}
\label{fig:top_charge_history}
\end{figure}

 Monte-Carlo histories of the topological charge $Q$ for $T = 220\mev$  are shown in Fig.~\ref{fig:top_charge_history}. While for $q B = 0$, $q B = 0.044 \gevq$ and $q B = 0.44 \gevq$ the Monte-Carlo histories look quite similar, for $q B = 0.22 \gevq$ we note that our Monte-Carlo process tends to spend more time in states with non-zero topological charge, which might imply some increase in topological susceptibility. To quantify this tendency we calculate the expectation value of the topological charge squared. We find that a small magnetic field ($q B = 0.044 \gevq$) causes almost no change in $\vev{Q^2}$ -- at $q B = 0$ we get $\vev{Q^2} = 0.264$, for $q B = 0.044\gevq$ this became  $\vev{Q^2} = 0.374$. At larger magnetic field ($q B = 0.22 \gevq$) there is a significant increase to $\vev{Q^2} = 0.870$, and then at $q B = 0.44 \gevq$ - the decrease almost to zero, $\vev{Q^2} = 0.02$. Unfortunately, due to very large autocorrelation time of the topological charge it is hardly possible to reliably estimate the errors of these numbers. It is interesting to note that such a behavior of $\vev{Q^2}$ parallels the non-monotonous dependence of the chiral condensate on magnetic field strength (see Fig.~\ref{fig:condensate_average}). However, with the present level of statistical uncertainties we cannot make reach quantitative conclusions on the dependence of $\vev{Q^2}$ on the magnetic field strength. It would be safer to say that our data just rule out any unexpectedly fast growth or decrease of $\vev{Q^2}$ at $q B \neq 0$. Finally, we note that in the deconfinement regime ($T = 280 \mev$) we did not see any topological fluctuations for any value of the magnetic field.

\section{Conclusions}
\label{sec:conclusions}

 In this paper we have presented the results of a lattice study of two-flavor lattice QCD with dynamical overlap fermions in an external magnetic field up to $0.68\gevq$ and with a pion mass around $500\mev$. Due to the use of special Hybrid Monte-Carlo algorithms developed for overlap fermions in \cite{Arnold:2003sx, Cundy:2004pza, Cundy:06, Cundy:09:1, Cundy:09:2, Cundy:11:2} we were able to perform fully first-principle simulations without any restriction of topological charge fluctuations and to generate several hundreds of configurations for each set of parameters.

 We have considered the dependence of the chiral condensate and the Polyakov loop on the magnetic field at two fixed lattice spacings ($a = 0.15 \fm$ and $a = 0.12 \fm$), which correspond to the temperatures $T = 220 \mev$ and $T = 280 \mev$ for the $16^3 \times 6$ lattice. The first value is likely to be very close to the deconfinement transition and the second value seems to be already in the deconfinement regime. Our results support the inverse magnetic catalysis scenario in which the deconfinement temperature decreases with increasing magnetic field.

 Finally, it is interesting to note that in the previous works \cite{Endrodi:12:jhep, Endrodi:12:prd, D'Elia:10} with staggered fermions inverse magnetic catalysis was observed only for sufficiently small pion masses. In contrast, in our simulations with chiral lattice fermions the pion mass is quite large (and is comparable to the value used in the seminal work \cite{D'Elia:10}), but nevertheless we find clear signatures of inverse magnetic catalysis. This observation suggests that good chiral properties seem to strengthen inverse magnetic catalysis. One possible explanation of this fact is the possible relation between inverse magnetic catalysis and the instanton-induced interactions between quarks \cite{Chao:13, Yu:14:1}. Correct implementation of such type of interactions (t'Hooft vertex) with rooted staggered fermions has been questioned by M.~Creutz for a long time \cite{Creutz:07:1,Creutz:08:1,Creutz:08:2}. While recent studies \cite{Bernard:08:1,Bernard:08:2,Kronfeld:11:1} demonstrate that such interactions might be still reproduced in rooted staggered QCD sufficiently close to the continuum limit, it is reasonable that at finite lattice spacing chiral lattice fermions will be more advantageous for studying the effects mediated by topological objects.

\begin{acknowledgments}
 We thank F.~Bruckmann, G.~Endrodi, E.~Fraga and T.~Sulejmanpasic for fruitful discussions. We deeply regret the sudden death of M.~I.~Polikarpov who has also strongly influenced this work. Computations were performed on \texttt{iDataCool} cluster of the University of Regensburg, \texttt{rrcmpi-a} cluster (``Graphyn'') at ITEP in Moscow, JSSC RAS (Moscow) and the ``Lomonosov'' and ``Chebyshev'' supercomputers at the supercomputing center of Moscow State University. VB is supported by grants RFBR 11-02-01227-a and RFBR 13-02-01387-a. PB and OK are supported by the S.~Kowalevskaja award by Alexander von Humboldt Foundation. NC is supported by the BK21 program of the NRF (MEST), Republic of Korea and Basic Science Research Program through the National Research Foundation of Korea(NRF) funded by the Ministry of Education(2013057640). AS is supported by SFB/TRR-55.
\end{acknowledgments}


\begin{thebibliography}{88}
\expandafter\ifx\csname natexlab\endcsname\relax\def\natexlab#1{#1}\fi
\expandafter\ifx\csname bibnamefont\endcsname\relax
  \def\bibnamefont#1{#1}\fi
\expandafter\ifx\csname bibfnamefont\endcsname\relax
  \def\bibfnamefont#1{#1}\fi
\expandafter\ifx\csname citenamefont\endcsname\relax
  \def\citenamefont#1{#1}\fi
\expandafter\ifx\csname url\endcsname\relax
  \def\url#1{\texttt{#1}}\fi
\expandafter\ifx\csname urlprefix\endcsname\relax\def\urlprefix{URL }\fi
\providecommand{\bibinfo}[2]{#2}
\providecommand{\eprint}[2][]{\url{#2}}

\bibitem[{\citenamefont{Skokov et~al.}(2009)\citenamefont{Skokov, Illarionov,
  and Toneev}}]{Skokov:09}
\bibinfo{author}{\bibfnamefont{V.}~\bibnamefont{Skokov}},
  \bibinfo{author}{\bibfnamefont{A.}~\bibnamefont{Illarionov}},
  \bibnamefont{and} \bibinfo{author}{\bibfnamefont{V.}~\bibnamefont{Toneev}},
  \bibinfo{journal}{Int. J. Mod. Phys. A} \textbf{\bibinfo{volume}{24}},
  \bibinfo{pages}{5925} (\bibinfo{year}{2009}),
  \href{http://arxiv.org/abs/0907.1396}{ArXiv:0907.1396}.

\bibitem[{\citenamefont{McLerran and Skokov}(2013)}]{Skokov:13}
\bibinfo{author}{\bibfnamefont{L.}~\bibnamefont{McLerran}} \bibnamefont{and}
  \bibinfo{author}{\bibfnamefont{V.}~\bibnamefont{Skokov}},
  \emph{\bibinfo{title}{Comments about the electromagnetic field in heavy-ion
  collisions}} (\bibinfo{year}{2013}),
  \href{http://arxiv.org/abs/1305.0774}{ArXiv:1305.0774}.

\bibitem[{\citenamefont{Kharzeev
  et~al.}(2008{\natexlab{a}})\citenamefont{Kharzeev, McLerran, and
  Warringa}}]{Kharzeev:08:1}
\bibinfo{author}{\bibfnamefont{D.~E.} \bibnamefont{Kharzeev}},
  \bibinfo{author}{\bibfnamefont{L.~D.} \bibnamefont{McLerran}},
  \bibnamefont{and} \bibinfo{author}{\bibfnamefont{H.~J.}
  \bibnamefont{Warringa}}, \bibinfo{journal}{Nucl. Phys. A}
  \textbf{\bibinfo{volume}{803}}, \bibinfo{pages}{227}
  (\bibinfo{year}{2008}{\natexlab{a}}),
  \href{http://arxiv.org/abs/0711.0950}{ArXiv:0711.0950}.

\bibitem[{\citenamefont{Romatschke}(2010)}]{Romatschke:09:1}
\bibinfo{author}{\bibfnamefont{P.}~\bibnamefont{Romatschke}},
  \bibinfo{journal}{Int.J.Mod.Phys.E} \textbf{\bibinfo{volume}{19}},
  \bibinfo{pages}{1} (\bibinfo{year}{2010}),
  \href{http://arxiv.org/abs/0902.3663}{ArXiv:0902.3663}.

\bibitem[{\citenamefont{Son and Surowka}(2009)}]{Son:09:1}
\bibinfo{author}{\bibfnamefont{D.~T.} \bibnamefont{Son}} \bibnamefont{and}
  \bibinfo{author}{\bibfnamefont{P.}~\bibnamefont{Surowka}},
  \bibinfo{journal}{Phys. Rev. Lett.} \textbf{\bibinfo{volume}{103}},
  \bibinfo{pages}{191601} (\bibinfo{year}{2009}),
  \href{http://arxiv.org/abs/0906.5044}{ArXiv:0906.5044}.

\bibitem[{\citenamefont{Sadofyev and Isachenkov}(2011)}]{Sadofyev:10:1}
\bibinfo{author}{\bibfnamefont{A.~V.} \bibnamefont{Sadofyev}} \bibnamefont{and}
  \bibinfo{author}{\bibfnamefont{M.~V.} \bibnamefont{Isachenkov}},
  \bibinfo{journal}{Phys. Lett. B} \textbf{\bibinfo{volume}{697}},
  \bibinfo{pages}{404 } (\bibinfo{year}{2011}),
  \href{http://arxiv.org/abs/1010.1550}{ArXiv:1010.1550}.

\bibitem[{\citenamefont{Sadofyev et~al.}(2011)\citenamefont{Sadofyev,
  Shevchenko, and Zakharov}}]{Sadofyev:10:2}
\bibinfo{author}{\bibfnamefont{A.~V.} \bibnamefont{Sadofyev}},
  \bibinfo{author}{\bibfnamefont{V.~I.} \bibnamefont{Shevchenko}},
  \bibnamefont{and} \bibinfo{author}{\bibfnamefont{V.~I.}
  \bibnamefont{Zakharov}}, \bibinfo{journal}{Phys. Rev. D}
  \textbf{\bibinfo{volume}{83}}, \bibinfo{pages}{105025}
  (\bibinfo{year}{2011}), \href{http://arxiv.org/abs/1012.1958}{ArXiv:1012.1958}.

\bibitem[{\citenamefont{Zakharov}(2012)}]{Zakharov:12:1}
\bibinfo{author}{\bibfnamefont{V.~I.} \bibnamefont{Zakharov}},
  \emph{\bibinfo{title}{Chiral magnetic effect in hydrodynamic approximation}},
  \bibinfo{howpublished}{in Lect. Notes Phys. {Strongly interacting matter in
  magnetic fields} (Springer), edited by D. Kharzeev, K. Landsteiner, A.
  Schmitt, H.-U. Yee} (\bibinfo{year}{2012}),
  \href{http://arxiv.org/abs/1210.2186}{ArXiv:1210.2186}.

\bibitem[{\citenamefont{Jensen}(2012)}]{Jensen:12:1}
\bibinfo{author}{\bibfnamefont{K.}~\bibnamefont{Jensen}},
  \bibinfo{journal}{Phys. Rev. D} \textbf{\bibinfo{volume}{85}},
  \bibinfo{pages}{125017} (\bibinfo{year}{2012}),
  \href{http://arxiv.org/abs/1203.3599}{ArXiv:1203.3599}.

\bibitem[{\citenamefont{Jensen et~al.}(2012)\citenamefont{Jensen, Kaminski,
  Kovtun, Meyer, Ritz, and Yarom}}]{Jensen:12:2}
\bibinfo{author}{\bibfnamefont{K.}~\bibnamefont{Jensen}},
  \bibinfo{author}{\bibfnamefont{M.}~\bibnamefont{Kaminski}},
  \bibinfo{author}{\bibfnamefont{P.}~\bibnamefont{Kovtun}},
  \bibinfo{author}{\bibfnamefont{R.}~\bibnamefont{Meyer}},
  \bibinfo{author}{\bibfnamefont{A.}~\bibnamefont{Ritz}}, \bibnamefont{and}
  \bibinfo{author}{\bibfnamefont{A.}~\bibnamefont{Yarom}},
  \bibinfo{journal}{Phys. Rev. Lett.} \textbf{\bibinfo{volume}{109}},
  \bibinfo{pages}{101601} (\bibinfo{year}{2012}),
  \href{http://arxiv.org/abs/1203.3556}{ArXiv:1203.3556}.

\bibitem[{\citenamefont{Banerjee et~al.}(2012)\citenamefont{Banerjee,
  Bhattacharya, Bhattacharyya, Jain, Minwalla, and Sharma}}]{Banerjee:12:1}
\bibinfo{author}{\bibfnamefont{N.}~\bibnamefont{Banerjee}},
  \bibinfo{author}{\bibfnamefont{J.}~\bibnamefont{Bhattacharya}},
  \bibinfo{author}{\bibfnamefont{S.}~\bibnamefont{Bhattacharyya}},
  \bibinfo{author}{\bibfnamefont{S.}~\bibnamefont{Jain}},
  \bibinfo{author}{\bibfnamefont{S.}~\bibnamefont{Minwalla}}, \bibnamefont{and}
  \bibinfo{author}{\bibfnamefont{T.}~\bibnamefont{Sharma}},
  \bibinfo{journal}{JHEP} \textbf{\bibinfo{volume}{09}}, \bibinfo{pages}{46}
  (\bibinfo{year}{2012}), \href{http://arxiv.org/abs/1203.3544}{ArXiv:1203.3544}.

\bibitem[{\citenamefont{Buividovich}(2013)}]{Buividovich:13:6}
\bibinfo{author}{\bibfnamefont{P.~V.} \bibnamefont{Buividovich}},
  \bibinfo{journal}{PoS} \textbf{\bibinfo{volume}{LATTICE2013}},
  \bibinfo{pages}{179} (\bibinfo{year}{2013}),
  \href{http://arxiv.org/abs/1309.2850}{ArXiv:1309.2850}.

\bibitem[{\citenamefont{Buividovich}(2014)}]{Buividovich:13:8}
\bibinfo{author}{\bibfnamefont{P.~V.} \bibnamefont{Buividovich}},
  \bibinfo{journal}{Nucl. Phys. A} \textbf{\bibinfo{volume}{925}},
  \bibinfo{pages}{218 } (\bibinfo{year}{2014}),
  \href{http://arxiv.org/abs/1312.1843}{ArXiv:1312.1843}.

\bibitem[{\citenamefont{D'Elia et~al.}(2010)\citenamefont{D'Elia, Mukherjee,
  and Sanfilippo}}]{D'Elia:10}
\bibinfo{author}{\bibfnamefont{M.}~\bibnamefont{D'Elia}},
  \bibinfo{author}{\bibfnamefont{S.}~\bibnamefont{Mukherjee}},
  \bibnamefont{and}
  \bibinfo{author}{\bibfnamefont{F.}~\bibnamefont{Sanfilippo}},
  \bibinfo{journal}{Phys. Rev. D} \textbf{\bibinfo{volume}{82}},
  \bibinfo{pages}{051501} (\bibinfo{year}{2010}),
  \href{http://arxiv.org/abs/1005.5365}{ArXiv:1005.5365}.

\bibitem[{\citenamefont{Bali et~al.}(2012)\citenamefont{Bali, Bruckmann,
  Endrodi, Fodor, Katz, Krieg, Sch\"afer, and Szabo}}]{Endrodi:12:jhep}
\bibinfo{author}{\bibfnamefont{G.~S.} \bibnamefont{Bali}},
  \bibinfo{author}{\bibfnamefont{F.}~\bibnamefont{Bruckmann}},
  \bibinfo{author}{\bibfnamefont{G.}~\bibnamefont{Endrodi}},
  \bibinfo{author}{\bibfnamefont{Z.}~\bibnamefont{Fodor}},
  \bibinfo{author}{\bibfnamefont{S.~D.} \bibnamefont{Katz}},
  \bibinfo{author}{\bibfnamefont{S.}~\bibnamefont{Krieg}},
  \bibinfo{author}{\bibfnamefont{A.}~\bibnamefont{Sch\"afer}},
  \bibnamefont{and} \bibinfo{author}{\bibfnamefont{K.~K.} \bibnamefont{Szabo}},
  \bibinfo{journal}{JHEP} \textbf{\bibinfo{volume}{02}}, \bibinfo{pages}{044}
  (\bibinfo{year}{2012}), \href{http://arxiv.org/abs/1111.4956}{ArXiv:1111.4956}.

\bibitem[{\citenamefont{Bruckmann
  et~al.}(2013{\natexlab{a}})\citenamefont{Bruckmann, Endrodi, and
  Kovacs}}]{Endrodi:13:jhep}
\bibinfo{author}{\bibfnamefont{F.}~\bibnamefont{Bruckmann}},
  \bibinfo{author}{\bibfnamefont{G.}~\bibnamefont{Endrodi}}, \bibnamefont{and}
  \bibinfo{author}{\bibfnamefont{T.~G.} \bibnamefont{Kovacs}},
  \bibinfo{journal}{JHEP} \textbf{\bibinfo{volume}{04}}, \bibinfo{pages}{112}
  (\bibinfo{year}{2013}{\natexlab{a}}),
  \href{http://arxiv.org/abs/1303.3972}{ArXiv:1303.3972}.

\bibitem[{\citenamefont{Bali et~al.}(12)\citenamefont{Bali, Bruckmann, Endrodi,
  Fodor, Katz, and Sch\"afer}}]{Endrodi:12:prd}
\bibinfo{author}{\bibfnamefont{G.~S.} \bibnamefont{Bali}},
  \bibinfo{author}{\bibfnamefont{F.}~\bibnamefont{Bruckmann}},
  \bibinfo{author}{\bibfnamefont{G.}~\bibnamefont{Endrodi}},
  \bibinfo{author}{\bibfnamefont{Z.}~\bibnamefont{Fodor}},
  \bibinfo{author}{\bibfnamefont{S.~D.} \bibnamefont{Katz}}, \bibnamefont{and}
  \bibinfo{author}{\bibfnamefont{A.}~\bibnamefont{Sch\"afer}},
  \bibinfo{journal}{Phys. Rev. D} \textbf{\bibinfo{volume}{86}},
  \bibinfo{pages}{071502} (\bibinfo{year}{12}),
  \href{http://arxiv.org/abs/1206.4205}{ArXiv:1206.4205}.

\bibitem[{\citenamefont{Ilgenfritz et~al.}(2014)\citenamefont{Ilgenfritz,
  M\"{u}ller-Preussker, Petersson, and Schreiber}}]{Muller-Preussker:13}
\bibinfo{author}{\bibfnamefont{E.}~\bibnamefont{Ilgenfritz}},
  \bibinfo{author}{\bibfnamefont{M.}~\bibnamefont{M\"{u}ller-Preussker}},
  \bibinfo{author}{\bibfnamefont{B.}~\bibnamefont{Petersson}},
  \bibnamefont{and}
  \bibinfo{author}{\bibfnamefont{A.}~\bibnamefont{Schreiber}},
  \bibinfo{journal}{Phys. Rev. D} \textbf{\bibinfo{volume}{89}},
  \bibinfo{pages}{054512} (\bibinfo{year}{2014}),
  \href{http://arxiv.org/abs/1310.7876}{ArXiv:1310.7876}.

\bibitem[{\citenamefont{Shushpanov and Smilga}(1997)}]{Smilga:97:1}
\bibinfo{author}{\bibfnamefont{I.~A.} \bibnamefont{Shushpanov}}
  \bibnamefont{and} \bibinfo{author}{\bibfnamefont{A.~V.}
  \bibnamefont{Smilga}}, \bibinfo{journal}{Phys. Lett. B}
  \textbf{\bibinfo{volume}{402}}, \bibinfo{pages}{351} (\bibinfo{year}{1997}),
  \href{http://arxiv.org/abs/hep-ph/9703201}{ArXiv:hep-ph/9703201}.

\bibitem[{\citenamefont{Cohen et~al.}(2007)\citenamefont{Cohen, McGady, and
  Werbos}}]{Cohen:07}
\bibinfo{author}{\bibfnamefont{T.~D.} \bibnamefont{Cohen}},
  \bibinfo{author}{\bibfnamefont{D.~A.} \bibnamefont{McGady}},
  \bibnamefont{and} \bibinfo{author}{\bibfnamefont{E.~S.}
  \bibnamefont{Werbos}}, \bibinfo{journal}{Phys. Rev. C}
  \textbf{\bibinfo{volume}{76}}, \bibinfo{pages}{055201}
  (\bibinfo{year}{2007}), \href{http://arxiv.org/abs/0706.3208}{ArXiv:0706.3208}.

\bibitem[{\citenamefont{Agasian}(2001)}]{Agasian:01}
\bibinfo{author}{\bibfnamefont{N.~O.} \bibnamefont{Agasian}},
  \bibinfo{journal}{Phys.Atom.Nucl.} \textbf{\bibinfo{volume}{64}},
  \bibinfo{pages}{554} (\bibinfo{year}{2001}),
  \href{http://arxiv.org/abs/hep-ph/0112341}{ArXiv:hep-ph/0112341}.

\bibitem[{\citenamefont{Andersen}(2012)}]{Andersen:12:1}
\bibinfo{author}{\bibfnamefont{J.~O.} \bibnamefont{Andersen}},
  \bibinfo{journal}{JHEP} \textbf{\bibinfo{volume}{1210}}, \bibinfo{pages}{005}
  (\bibinfo{year}{2012}), \href{http://arxiv.org/abs/1205.6978}{ArXiv:1205.6978}.

\bibitem[{\citenamefont{Andersen and Tranberg}(2012)}]{Andersen:12:2}
\bibinfo{author}{\bibfnamefont{J.~O.} \bibnamefont{Andersen}} \bibnamefont{and}
  \bibinfo{author}{\bibfnamefont{A.}~\bibnamefont{Tranberg}},
  \bibinfo{journal}{JHEP} \textbf{\bibinfo{volume}{1208}}, \bibinfo{pages}{002}
  (\bibinfo{year}{2012}), \href{http://arxiv.org/abs/1204.3360}{ArXiv:1204.3360}.

\bibitem[{\citenamefont{Andersen et~al.}(2013)\citenamefont{Andersen, Naylor,
  and Tranberg}}]{Andersen:13}
\bibinfo{author}{\bibfnamefont{J.~O.} \bibnamefont{Andersen}},
  \bibinfo{author}{\bibfnamefont{W.~R.} \bibnamefont{Naylor}},
  \bibnamefont{and} \bibinfo{author}{\bibfnamefont{A.}~\bibnamefont{Tranberg}}
  (\bibinfo{year}{2013}), \href{http://arxiv.org/abs/1311.2093}{ArXiv:1311.2093}.

\bibitem[{\citenamefont{Scherer and Gies}(2012)}]{Scherer:12}
\bibinfo{author}{\bibfnamefont{D.~D.} \bibnamefont{Scherer}} \bibnamefont{and}
  \bibinfo{author}{\bibfnamefont{H.}~\bibnamefont{Gies}},
  \bibinfo{journal}{Phys. Rev. B} \textbf{\bibinfo{volume}{85}},
  \bibinfo{pages}{195417} (\bibinfo{year}{2012}),
  \href{http://arxiv.org/abs/1201.3746}{ArXiv:1201.3746}.

\bibitem[{\citenamefont{Kanemura et~al.}(1998)\citenamefont{Kanemura, Sato, and
  Tochimura}}]{Sato:98}
\bibinfo{author}{\bibfnamefont{S.}~\bibnamefont{Kanemura}},
  \bibinfo{author}{\bibfnamefont{H.-T.} \bibnamefont{Sato}}, \bibnamefont{and}
  \bibinfo{author}{\bibfnamefont{H.}~\bibnamefont{Tochimura}},
  \bibinfo{journal}{Nucl. Phys. B} \textbf{\bibinfo{volume}{517}},
  \bibinfo{pages}{567} (\bibinfo{year}{1998}),
  \href{http://arxiv.org/abs/hep-ph/9707285}{ArXiv:hep-ph/9707285}.

\bibitem[{\citenamefont{Johnson and Kundu}(2008)}]{Johnson:08}
\bibinfo{author}{\bibfnamefont{C.~V.} \bibnamefont{Johnson}} \bibnamefont{and}
  \bibinfo{author}{\bibfnamefont{A.}~\bibnamefont{Kundu}},
  \bibinfo{journal}{JHEP} \textbf{\bibinfo{volume}{0812}}, \bibinfo{pages}{053}
  (\bibinfo{year}{2008}), \href{http://arxiv.org/abs/0803.0038}{ArXiv:0803.0038}.

\bibitem[{\citenamefont{Fraga and Mizher}(2008)}]{Fraga:08}
\bibinfo{author}{\bibfnamefont{E.~S.} \bibnamefont{Fraga}} \bibnamefont{and}
  \bibinfo{author}{\bibfnamefont{A.}~\bibnamefont{Mizher}},
  \bibinfo{journal}{Phys. Rev. D} \textbf{\bibinfo{volume}{78}},
  \bibinfo{pages}{025016} (\bibinfo{year}{2008}),
  \href{http://arxiv.org/abs/0804.1452}{ArXiv:0804.1452}.

\bibitem[{\citenamefont{Gatto and Ruggieri}(2011)}]{Ruggieri:2011}
\bibinfo{author}{\bibfnamefont{R.}~\bibnamefont{Gatto}} \bibnamefont{and}
  \bibinfo{author}{\bibfnamefont{M.}~\bibnamefont{Ruggieri}},
  \bibinfo{journal}{Phys. Rev. D} \textbf{\bibinfo{volume}{83}},
  \bibinfo{pages}{034016} (\bibinfo{year}{2011}),
  \href{http://arxiv.org/abs/1012.1291}{ArXiv:1012.1291}.

\bibitem[{\citenamefont{Gatto and Ruggieri}(2010)}]{Ruggieri:10}
\bibinfo{author}{\bibfnamefont{R.}~\bibnamefont{Gatto}} \bibnamefont{and}
  \bibinfo{author}{\bibfnamefont{M.}~\bibnamefont{Ruggieri}},
  \bibinfo{journal}{Phys. Rev. D} \textbf{\bibinfo{volume}{82}},
  \bibinfo{pages}{054027} (\bibinfo{year}{2010}),
  \href{http://arxiv.org/abs/1007.0790}{ArXiv:1007.0790}.

\bibitem[{\citenamefont{Kashiwa}(2011)}]{Kashiwa:11}
\bibinfo{author}{\bibfnamefont{K.}~\bibnamefont{Kashiwa}},
  \bibinfo{journal}{Phys. Rev. D} \textbf{\bibinfo{volume}{83}},
  \bibinfo{pages}{117901} (\bibinfo{year}{2011}),
  \href{http://arxiv.org/abs/1104.5167}{ArXiv:1104.5167}.

\bibitem[{\citenamefont{Endrodi}(2013)}]{Endrodi:2013cs}
\bibinfo{author}{\bibfnamefont{G.}~\bibnamefont{Endrodi}},
  \bibinfo{journal}{JHEP} \textbf{\bibinfo{volume}{04}}, \bibinfo{pages}{023}
  (\bibinfo{year}{2013}), \href{http://arxiv.org/abs/1301.1307}{ArXiv:1301.1307}.

\bibitem[{\citenamefont{Blaizot et~al.}(2013)\citenamefont{Blaizot, Fraga, and
  Palhares}}]{Fraga:13plb}
\bibinfo{author}{\bibfnamefont{J.}~\bibnamefont{Blaizot}},
  \bibinfo{author}{\bibfnamefont{E.~S.} \bibnamefont{Fraga}}, \bibnamefont{and}
  \bibinfo{author}{\bibfnamefont{L.~F.} \bibnamefont{Palhares}},
  \bibinfo{journal}{Phys. Lett. B} \textbf{\bibinfo{volume}{722}},
  \bibinfo{pages}{167} (\bibinfo{year}{2013}),
  \href{http://arxiv.org/abs/1211.6412}{ArXiv:1211.6412}.

\bibitem[{\citenamefont{Orlovsky and Simonov}(2014)}]{Simonov:14}
\bibinfo{author}{\bibfnamefont{V.~D.} \bibnamefont{Orlovsky}} \bibnamefont{and}
  \bibinfo{author}{\bibfnamefont{Y.~A.} \bibnamefont{Simonov}},
  \bibinfo{journal}{Phys. Rev. D} \textbf{\bibinfo{volume}{89}},
  \bibinfo{pages}{054012} (\bibinfo{year}{2014}),
  \href{http://arxiv.org/abs/1311.1087}{ArXiv:1311.1087}.

\bibitem[{\citenamefont{Bruckmann
  et~al.}(2013{\natexlab{b}})\citenamefont{Bruckmann, Endrodi, and
  Kovacs}}]{Endrodi:13:new}
\bibinfo{author}{\bibfnamefont{F.}~\bibnamefont{Bruckmann}},
  \bibinfo{author}{\bibfnamefont{G.}~\bibnamefont{Endrodi}}, \bibnamefont{and}
  \bibinfo{author}{\bibfnamefont{T.~G.} \bibnamefont{Kovacs}},
  \emph{\bibinfo{title}{Inverse magnetic catalysis in {QCD}}}
  (\bibinfo{year}{2013}{\natexlab{b}}),
  \href{http://arxiv.org/abs/1311.3178}{ArXiv:1311.3178}.

\bibitem[{\citenamefont{Fraga et~al.}(2013)\citenamefont{Fraga, Noronha, and
  Palhares}}]{Fraga:13prd}
\bibinfo{author}{\bibfnamefont{E.~S.} \bibnamefont{Fraga}},
  \bibinfo{author}{\bibfnamefont{J.}~\bibnamefont{Noronha}}, \bibnamefont{and}
  \bibinfo{author}{\bibfnamefont{L.~F.} \bibnamefont{Palhares}},
  \bibinfo{journal}{Phys. Rev. D} \textbf{\bibinfo{volume}{87}},
  \bibinfo{pages}{114014} (\bibinfo{year}{2013}),
  \href{http://arxiv.org/abs/1207.7094}{ArXiv:1207.7094}.

\bibitem[{\citenamefont{Ballon-Bayona}(2013)}]{Ballon-Bayona:13}
\bibinfo{author}{\bibfnamefont{A.}~\bibnamefont{Ballon-Bayona}},
  \bibinfo{journal}{JHEP} \textbf{\bibinfo{volume}{1311}}, \bibinfo{pages}{168}
  (\bibinfo{year}{2013}), \href{http://arxiv.org/abs/1307.6498}{ArXiv:1307.6498}.

\bibitem[{\citenamefont{Fraga and Palhares}(2012)}]{Fraga:12}
\bibinfo{author}{\bibfnamefont{E.~S.} \bibnamefont{Fraga}} \bibnamefont{and}
  \bibinfo{author}{\bibfnamefont{L.~F.} \bibnamefont{Palhares}},
  \bibinfo{journal}{Phys. Rev. D} \textbf{\bibinfo{volume}{86}},
  \bibinfo{pages}{016008} (\bibinfo{year}{2012}),
  \href{http://arxiv.org/abs/1201.5881}{ArXiv:1201.5881}.

\bibitem[{\citenamefont{Ferreira
  et~al.}(2014{\natexlab{a}})\citenamefont{Ferreira, Costa, Lourenco,
  Frederico, and Providencia}}]{Ferreira:14}
\bibinfo{author}{\bibfnamefont{M.}~\bibnamefont{Ferreira}},
  \bibinfo{author}{\bibfnamefont{P.}~\bibnamefont{Costa}},
  \bibinfo{author}{\bibfnamefont{O.}~\bibnamefont{Lourenco}},
  \bibinfo{author}{\bibfnamefont{T.}~\bibnamefont{Frederico}},
  \bibnamefont{and}
  \bibinfo{author}{\bibfnamefont{C.}~\bibnamefont{Providencia}},
  \bibinfo{journal}{Phys. Rev. D} \textbf{\bibinfo{volume}{89}},
  \bibinfo{pages}{116011} (\bibinfo{year}{2014}{\natexlab{a}}),
  \href{http://arxiv.org/abs/1404.5577}{ArXiv:1404.5577}.

\bibitem[{\citenamefont{Ferreira
  et~al.}(2014{\natexlab{b}})\citenamefont{Ferreira, Costa, Menezes,
  Providencia, and Scoccola}}]{Ferreira:14prd}
\bibinfo{author}{\bibfnamefont{M.}~\bibnamefont{Ferreira}},
  \bibinfo{author}{\bibfnamefont{P.}~\bibnamefont{Costa}},
  \bibinfo{author}{\bibfnamefont{D.~P.} \bibnamefont{Menezes}},
  \bibinfo{author}{\bibfnamefont{C.}~\bibnamefont{Providencia}},
  \bibnamefont{and} \bibinfo{author}{\bibfnamefont{N.}~\bibnamefont{Scoccola}},
  \bibinfo{journal}{Phys. Rev. D} \textbf{\bibinfo{volume}{89}},
  \bibinfo{pages}{016002} (\bibinfo{year}{2014}{\natexlab{b}}),
  \href{http://arxiv.org/abs/1305.4751}{ArXiv:1305.4751}.

\bibitem[{\citenamefont{Chao et~al.}(2013)\citenamefont{Chao, Chu, and
  Huang}}]{Chao:13}
\bibinfo{author}{\bibfnamefont{J.}~\bibnamefont{Chao}},
  \bibinfo{author}{\bibfnamefont{P.}~\bibnamefont{Chu}}, \bibnamefont{and}
  \bibinfo{author}{\bibfnamefont{M.}~\bibnamefont{Huang}},
  \bibinfo{journal}{Phys. Rev. D} \textbf{\bibinfo{volume}{88}},
  \bibinfo{pages}{054009} (\bibinfo{year}{2013}),
  \href{http://arxiv.org/abs/1305.1100}{ArXiv:1305.1100}.

\bibitem[{\citenamefont{Yu et~al.}(2014)\citenamefont{Yu, Liu, and
  Huang}}]{Yu:14:1}
\bibinfo{author}{\bibfnamefont{L.}~\bibnamefont{Yu}},
  \bibinfo{author}{\bibfnamefont{H.}~\bibnamefont{Liu}}, \bibnamefont{and}
  \bibinfo{author}{\bibfnamefont{M.}~\bibnamefont{Huang}},
  \emph{\bibinfo{title}{Spontaneous generation of local {CP} violation and
  inverse magnetic catalysis}} (\bibinfo{year}{2014}),
  \href{http://arxiv.org/abs/1404.6969}{ArXiv:1404.6969}.

\bibitem[{\citenamefont{Ruggieri}(2011)}]{Ruggieri:11:1}
\bibinfo{author}{\bibfnamefont{M.}~\bibnamefont{Ruggieri}},
  \bibinfo{journal}{Phys. Rev. D} \textbf{\bibinfo{volume}{84}},
  \bibinfo{pages}{014011} (\bibinfo{year}{2011}),
  \href{http://arxiv.org/abs/1103.6186}{ArXiv:1103.6186}.

\bibitem[{\citenamefont{Gatto and Ruggieri}(2012)}]{Gatto:12:1}
\bibinfo{author}{\bibfnamefont{R.}~\bibnamefont{Gatto}} \bibnamefont{and}
  \bibinfo{author}{\bibfnamefont{M.}~\bibnamefont{Ruggieri}},
  \bibinfo{journal}{Phys. Rev. D} \textbf{\bibinfo{volume}{85}},
  \bibinfo{pages}{054013} (\bibinfo{year}{2012}),
  \href{http://arxiv.org/abs/1110.4904}{ArXiv:1110.4904}.

\bibitem[{\citenamefont{Chernodub and Nedelin}(2011)}]{Nedelin:11:1}
\bibinfo{author}{\bibfnamefont{M.~N.} \bibnamefont{Chernodub}}
  \bibnamefont{and} \bibinfo{author}{\bibfnamefont{A.~S.}
  \bibnamefont{Nedelin}}, \bibinfo{journal}{Phys. Rev. D}
  \textbf{\bibinfo{volume}{83}}, \bibinfo{pages}{105008}
  (\bibinfo{year}{2011}), \href{http://arxiv.org/abs/1102.0188}{ArXiv:1102.0188}.

\bibitem[{\citenamefont{Schaefer et~al.}(1995)\citenamefont{Schaefer, Shuryak,
  and Verbaarschot}}]{Shuryak:94:1}
\bibinfo{author}{\bibfnamefont{T.}~\bibnamefont{Schaefer}},
  \bibinfo{author}{\bibfnamefont{E.}~\bibnamefont{Shuryak}}, \bibnamefont{and}
  \bibinfo{author}{\bibfnamefont{J.}~\bibnamefont{Verbaarschot}},
  \bibinfo{journal}{Phys. Rev. D} \textbf{\bibinfo{volume}{51}},
  \bibinfo{pages}{1267} (\bibinfo{year}{1995}),
  \href{http://arxiv.org/abs/hep-ph/9406210}{ArXiv:hep-ph/9406210}.

\bibitem[{\citenamefont{Ilgenfritz and Shuryak}(1994)}]{Shuryak:94:2}
\bibinfo{author}{\bibfnamefont{E.}~\bibnamefont{Ilgenfritz}} \bibnamefont{and}
  \bibinfo{author}{\bibfnamefont{E.~V.} \bibnamefont{Shuryak}},
  \bibinfo{journal}{Phys. Lett. B} \textbf{\bibinfo{volume}{325}},
  \bibinfo{pages}{263 } (\bibinfo{year}{1994}),
  \href{http://arxiv.org/abs/hep-ph/9401285}{ArXiv:hep-ph/9401285}.

\bibitem[{\citenamefont{Creutz}(2007)}]{Creutz:07:1}
\bibinfo{author}{\bibfnamefont{M.}~\bibnamefont{Creutz}},
  \bibinfo{journal}{Phys. Lett. B} \textbf{\bibinfo{volume}{649}},
  \bibinfo{pages}{230 } (\bibinfo{year}{2007}),
  \href{http://arxiv.org/abs/hep-lat/0701018}{ArXiv:hep-lat/0701018}.

\bibitem[{\citenamefont{Creutz}(2008{\natexlab{a}})}]{Creutz:08:1}
\bibinfo{author}{\bibfnamefont{M.}~\bibnamefont{Creutz}},
  \bibinfo{journal}{Ann. Phys.} \textbf{\bibinfo{volume}{323}},
  \bibinfo{pages}{2349 } (\bibinfo{year}{2008}{\natexlab{a}}),
  \href{http://arxiv.org/abs/0711.2640}{ArXiv:0711.2640}.

\bibitem[{\citenamefont{Creutz}(2008{\natexlab{b}})}]{Creutz:08:2}
\bibinfo{author}{\bibfnamefont{M.}~\bibnamefont{Creutz}},
  \bibinfo{journal}{Phys. Rev. D} \textbf{\bibinfo{volume}{78}},
  \bibinfo{pages}{078501} (\bibinfo{year}{2008}{\natexlab{b}}),
  \href{http://arxiv.org/abs/0805.1350}{ArXiv:0805.1350}.

\bibitem[{\citenamefont{Bernard
  et~al.}(2008{\natexlab{a}})\citenamefont{Bernard, Golterman, Shamir, and
  Sharpe}}]{Bernard:08:1}
\bibinfo{author}{\bibfnamefont{C.}~\bibnamefont{Bernard}},
  \bibinfo{author}{\bibfnamefont{M.}~\bibnamefont{Golterman}},
  \bibinfo{author}{\bibfnamefont{Y.}~\bibnamefont{Shamir}}, \bibnamefont{and}
  \bibinfo{author}{\bibfnamefont{S.~R.} \bibnamefont{Sharpe}},
  \bibinfo{journal}{Phys. Rev. D} \textbf{\bibinfo{volume}{77}},
  \bibinfo{pages}{114504} (\bibinfo{year}{2008}{\natexlab{a}}),
  \href{http://arxiv.org/abs/0711.0696}{ArXiv:0711.0696}.

\bibitem[{\citenamefont{Bernard
  et~al.}(2008{\natexlab{b}})\citenamefont{Bernard, Golterman, Shamir, and
  Sharpe}}]{Bernard:08:2}
\bibinfo{author}{\bibfnamefont{C.}~\bibnamefont{Bernard}},
  \bibinfo{author}{\bibfnamefont{M.}~\bibnamefont{Golterman}},
  \bibinfo{author}{\bibfnamefont{Y.}~\bibnamefont{Shamir}}, \bibnamefont{and}
  \bibinfo{author}{\bibfnamefont{S.~R.} \bibnamefont{Sharpe}},
  \bibinfo{journal}{Phys. Rev. D} \textbf{\bibinfo{volume}{78}},
  \bibinfo{pages}{078502} (\bibinfo{year}{2008}{\natexlab{b}}),
  \href{http://arxiv.org/abs/0808.2056}{ArXiv:0808.2056}.

\bibitem[{\citenamefont{Donald et~al.}(2011)\citenamefont{Donald, Davies,
  Follana, and Kronfeld}}]{Kronfeld:11:1}
\bibinfo{author}{\bibfnamefont{G.~C.} \bibnamefont{Donald}},
  \bibinfo{author}{\bibfnamefont{C.~T.~H.} \bibnamefont{Davies}},
  \bibinfo{author}{\bibfnamefont{E.}~\bibnamefont{Follana}}, \bibnamefont{and}
  \bibinfo{author}{\bibfnamefont{A.~S.} \bibnamefont{Kronfeld}},
  \bibinfo{journal}{Phys. Rev. D} \textbf{\bibinfo{volume}{84}},
  \bibinfo{pages}{054504} (\bibinfo{year}{2011}),
  \href{http://arxiv.org/abs/1106.2412}{ArXiv:1106.2412}.

\bibitem[{\citenamefont{Kharzeev and Warringa}(2009)}]{Kharzeev:09:1}
\bibinfo{author}{\bibfnamefont{D.~E.} \bibnamefont{Kharzeev}} \bibnamefont{and}
  \bibinfo{author}{\bibfnamefont{H.~J.} \bibnamefont{Warringa}},
  \bibinfo{journal}{Phys. Rev. D} \textbf{\bibinfo{volume}{80}},
  \bibinfo{pages}{034028} (\bibinfo{year}{2009}),
  \href{http://arxiv.org/abs/0907.5007}{ArXiv:0907.5007}.

\bibitem[{\citenamefont{Newman and Son}(2006)}]{Son:06:2}
\bibinfo{author}{\bibfnamefont{G.~M.} \bibnamefont{Newman}} \bibnamefont{and}
  \bibinfo{author}{\bibfnamefont{D.~T.} \bibnamefont{Son}},
  \bibinfo{journal}{Phys. Rev. D} \textbf{\bibinfo{volume}{73}},
  \bibinfo{pages}{045006} (\bibinfo{year}{2006}),
  \href{http://arxiv.org/abs/hep-ph/0510049}{ArXiv:hep-ph/0510049}.

\bibitem[{\citenamefont{Metlitski and Zhitnitsky}(2005)}]{Metlitski:05:1}
\bibinfo{author}{\bibfnamefont{M.~A.} \bibnamefont{Metlitski}}
  \bibnamefont{and} \bibinfo{author}{\bibfnamefont{A.~R.}
  \bibnamefont{Zhitnitsky}}, \bibinfo{journal}{Phys. Rev. D}
  \textbf{\bibinfo{volume}{72}}, \bibinfo{pages}{045011}
  (\bibinfo{year}{2005}), \href{http://arxiv.org/abs/hep-ph/0505072}{ArXiv:hep-ph/0505072}.

\bibitem[{\citenamefont{Jensen et~al.}(2013)\citenamefont{Jensen, Kovtun, and
  Ritz}}]{Jensen:13:1}
\bibinfo{author}{\bibfnamefont{K.}~\bibnamefont{Jensen}},
  \bibinfo{author}{\bibfnamefont{P.}~\bibnamefont{Kovtun}}, \bibnamefont{and}
  \bibinfo{author}{\bibfnamefont{A.}~\bibnamefont{Ritz}},
  \bibinfo{journal}{JHEP} \textbf{\bibinfo{volume}{1310}}, \bibinfo{pages}{186}
  (\bibinfo{year}{2013}), \href{http://arxiv.org/abs/1307.3234}{ArXiv:1307.3234}.

\bibitem[{\citenamefont{Gursoy and Jansen}(2014)}]{Gursoy:14:1}
\bibinfo{author}{\bibfnamefont{U.}~\bibnamefont{Gursoy}} \bibnamefont{and}
  \bibinfo{author}{\bibfnamefont{A.}~\bibnamefont{Jansen}},
  \emph{\bibinfo{title}{(non)renormalization of anomalous conductivities and
  holography}} (\bibinfo{year}{2014}),
  \href{http://arxiv.org/abs/1407.3282}{ArXiv:1407.3282}.

\bibitem[{\citenamefont{Arnold et~al.}(2003)\citenamefont{Arnold, Cundy, {van
  den Eshof}, Frommer, and Krieg}}]{Arnold:2003sx}
\bibinfo{author}{\bibfnamefont{G.}~\bibnamefont{Arnold}},
  \bibinfo{author}{\bibfnamefont{N.}~\bibnamefont{Cundy}},
  \bibinfo{author}{\bibfnamefont{J.}~\bibnamefont{{van den Eshof}}},
  \bibinfo{author}{\bibfnamefont{A.}~\bibnamefont{Frommer}}, \bibnamefont{and}
  \bibinfo{author}{\bibfnamefont{S.}~\bibnamefont{Krieg}},
  \emph{\bibinfo{title}{Numerical methods for the {QCD} overlap operator. 2.
  {O}ptimal {K}rylov subspace methods}} (\bibinfo{year}{2003}),
  \href{http://arxiv.org/abs/hep-lat/0311025}{ArXiv:hep-lat/0311025}.

\bibitem[{\citenamefont{Cundy et~al.}(2005)\citenamefont{Cundy, {van den
  Eshof}, Frommer, Krieg, and Lippert}}]{Cundy:2004pza}
\bibinfo{author}{\bibfnamefont{N.}~\bibnamefont{Cundy}},
  \bibinfo{author}{\bibfnamefont{J.}~\bibnamefont{{van den Eshof}}},
  \bibinfo{author}{\bibfnamefont{A.}~\bibnamefont{Frommer}},
  \bibinfo{author}{\bibfnamefont{S.}~\bibnamefont{Krieg}}, \bibnamefont{and}
  \bibinfo{author}{\bibfnamefont{T.}~\bibnamefont{Lippert}},
  \bibinfo{journal}{Comput.Phys.Commun.} \textbf{\bibinfo{volume}{165}},
  \bibinfo{pages}{221} (\bibinfo{year}{2005}),
  \href{http://arxiv.org/abs/hep-lat/0311025}{ArXiv:hep-lat/0311025}.

\bibitem[{\citenamefont{Cundy}(2006)}]{Cundy:06}
\bibinfo{author}{\bibfnamefont{N.}~\bibnamefont{Cundy}},
  \bibinfo{journal}{Nucl.Phys.Proc.Suppl.} \textbf{\bibinfo{volume}{153}},
  \bibinfo{pages}{54} (\bibinfo{year}{2006}),
  \href{http://arxiv.org/abs/hep-lat/0511047}{ArXiv:hep-lat/0511047}.

\bibitem[{\citenamefont{Cundy et~al.}(2009)\citenamefont{Cundy, Krieg, Arnold,
  Frommer, Lippert, and Schilling}}]{Cundy:09:1}
\bibinfo{author}{\bibfnamefont{N.}~\bibnamefont{Cundy}},
  \bibinfo{author}{\bibfnamefont{S.}~\bibnamefont{Krieg}},
  \bibinfo{author}{\bibfnamefont{G.}~\bibnamefont{Arnold}},
  \bibinfo{author}{\bibfnamefont{A.}~\bibnamefont{Frommer}},
  \bibinfo{author}{\bibfnamefont{T.}~\bibnamefont{Lippert}}, \bibnamefont{and}
  \bibinfo{author}{\bibfnamefont{K.}~\bibnamefont{Schilling}},
  \bibinfo{journal}{Comput.Phys.Commun.} \textbf{\bibinfo{volume}{180}},
  \bibinfo{pages}{26} (\bibinfo{year}{2009}),
  \href{http://arxiv.org/abs/hep-lat/0502007}{ArXiv:hep-lat/0502007}.

\bibitem[{\citenamefont{Cundy}(2009)}]{Cundy:09:2}
\bibinfo{author}{\bibfnamefont{N.}~\bibnamefont{Cundy}},
  \bibinfo{journal}{Comput.Phys.Commun} \textbf{\bibinfo{volume}{180}},
  \bibinfo{pages}{180} (\bibinfo{year}{2009}),
  \href{http://arxiv.org/abs/0706.1971}{ArXiv:0706.1971}.

\bibitem[{\citenamefont{Cundy and Lee}(2011)}]{Cundy:11:2}
\bibinfo{author}{\bibfnamefont{N.}~\bibnamefont{Cundy}} \bibnamefont{and}
  \bibinfo{author}{\bibfnamefont{W.}~\bibnamefont{Lee}},
  \emph{\bibinfo{title}{Modifying the molecular dynamics action to increase
  topological tunnelling rate for dynamical overlap fermions}}
  (\bibinfo{year}{2011}), \href{http://arxiv.org/abs/1110.1948}{ArXiv:1110.1948}.

\bibitem[{\citenamefont{Bali et~al.}(2013)\citenamefont{Bali, Bruckmann,
  Constantinou, Costa, Endrodi, Katz, Panagopoulos, and Schafer}}]{Bali:13:1}
\bibinfo{author}{\bibfnamefont{G.~S.} \bibnamefont{Bali}},
  \bibinfo{author}{\bibfnamefont{F.}~\bibnamefont{Bruckmann}},
  \bibinfo{author}{\bibfnamefont{M.}~\bibnamefont{Constantinou}},
  \bibinfo{author}{\bibfnamefont{M.}~\bibnamefont{Costa}},
  \bibinfo{author}{\bibfnamefont{G.}~\bibnamefont{Endrodi}},
  \bibinfo{author}{\bibfnamefont{S.~D.} \bibnamefont{Katz}},
  \bibinfo{author}{\bibfnamefont{H.}~\bibnamefont{Panagopoulos}},
  \bibnamefont{and} \bibinfo{author}{\bibfnamefont{A.}~\bibnamefont{Schafer}},
  \bibinfo{journal}{PoS} \textbf{\bibinfo{volume}{LATTICE2013}},
  \bibinfo{pages}{458} (\bibinfo{year}{2013}),
  \href{http://arxiv.org/abs/1311.3519}{ArXiv:1311.3519}.

\bibitem[{\citenamefont{Bali et~al.}(2014)\citenamefont{Bali, Bruckmann,
  Endrodi, Katz, and Schafer}}]{Bali:14:1}
\bibinfo{author}{\bibfnamefont{G.~S.} \bibnamefont{Bali}},
  \bibinfo{author}{\bibfnamefont{F.}~\bibnamefont{Bruckmann}},
  \bibinfo{author}{\bibfnamefont{G.}~\bibnamefont{Endrodi}},
  \bibinfo{author}{\bibfnamefont{S.~D.} \bibnamefont{Katz}}, \bibnamefont{and}
  \bibinfo{author}{\bibfnamefont{A.}~\bibnamefont{Schafer}},
  \emph{\bibinfo{title}{The {QCD} equation of state in background magnetic
  fields}} (\bibinfo{year}{2014}),
  \href{http://arxiv.org/abs/1406.0269}{ArXiv:1406.0269}.

\bibitem[{\citenamefont{Moran and Leinweber}(2008)}]{Moran:2008ra}
\bibinfo{author}{\bibfnamefont{P.~J.} \bibnamefont{Moran}} \bibnamefont{and}
  \bibinfo{author}{\bibfnamefont{D.~B.} \bibnamefont{Leinweber}},
  \bibinfo{journal}{Phys. Rev. D} \textbf{\bibinfo{volume}{77}},
  \bibinfo{pages}{094501} (\bibinfo{year}{2008}),
  \href{http://arxiv.org/abs/0801.1165}{ArXiv:0801.1165}.

\bibitem[{\citenamefont{Morningstar and Peardon}(2004)}]{Morningstar:2003gk}
\bibinfo{author}{\bibfnamefont{C.}~\bibnamefont{Morningstar}} \bibnamefont{and}
  \bibinfo{author}{\bibfnamefont{M.}~\bibnamefont{Peardon}},
  \bibinfo{journal}{Phys. Rev. D} \textbf{\bibinfo{volume}{69}},
  \bibinfo{pages}{054501} (\bibinfo{year}{2004}),
  \href{http://arxiv.org/abs/hep-lat/0311018}{ArXiv:hep-lat/0311018}.

\bibitem[{\citenamefont{L{\"u}scher and Weisz}(1985)}]{TILW}
\bibinfo{author}{\bibfnamefont{M.}~\bibnamefont{L{\"u}scher}} \bibnamefont{and}
  \bibinfo{author}{\bibfnamefont{P.}~\bibnamefont{Weisz}},
  \href{http://dx.doi.org/10.1007/BF01206178}{  \bibinfo{journal}{Commun. Math. Phys.} \textbf{\bibinfo{volume}{97}},
  \bibinfo{pages}{59} (\bibinfo{year}{1985})}.

\bibitem[{\citenamefont{Lepage and Mackenzie}(1993)}]{TILW2}
\bibinfo{author}{\bibfnamefont{G.~P.} \bibnamefont{Lepage}} \bibnamefont{and}
  \bibinfo{author}{\bibfnamefont{P.~B.} \bibnamefont{Mackenzie}},
  \bibinfo{journal}{Phys. Rev. D} \textbf{\bibinfo{volume}{48}},
  \bibinfo{pages}{2250} (\bibinfo{year}{1993}),
  \href{http://arxiv.org/abs/hep-lat/9209022}{ArXiv:hep-lat/9209022}.

\bibitem[{\citenamefont{Snippe}(1997)}]{TILW4}
\bibinfo{author}{\bibfnamefont{J.}~\bibnamefont{Snippe}},
  \href{http://dx.doi.org/10.1016/S0550-3213(97)00270-8}{  \bibinfo{journal}{Nucl. Phys. B} \textbf{\bibinfo{volume}{498}},
  \bibinfo{pages}{347} (\bibinfo{year}{1997})}.

\bibitem[{\citenamefont{Curci et~al.}(1983)\citenamefont{Curci, Menotti, and
  Paffuti}}]{TILW5}
\bibinfo{author}{\bibfnamefont{G.}~\bibnamefont{Curci}},
  \bibinfo{author}{\bibfnamefont{P.}~\bibnamefont{Menotti}}, \bibnamefont{and}
  \bibinfo{author}{\bibfnamefont{G.}~\bibnamefont{Paffuti}},
  \href{http://dx.doi.org/10.1016/0370-2693(83)91043-2}{  \bibinfo{journal}{Phys. Lett. B} \textbf{\bibinfo{volume}{130}},
  \bibinfo{pages}{205} (\bibinfo{year}{1983})}.

\bibitem[{\citenamefont{Luschevskaya}(2014)}]{Luschevskaya:12}
\bibinfo{author}{\bibfnamefont{E.~V.} \bibnamefont{Luschevskaya}},
  \bibinfo{journal}{Nucl. Phys. B} \textbf{\bibinfo{volume}{884}},
  \bibinfo{pages}{1} (\bibinfo{year}{2014}),
  \href{http://arxiv.org/abs/1203.5699}{ArXiv:1203.5699}.

\bibitem[{\citenamefont{Sommer}(1994)}]{Sommer:1993ce}
\bibinfo{author}{\bibfnamefont{R.}~\bibnamefont{Sommer}},
  \bibinfo{journal}{Nucl. Phys. B} \textbf{\bibinfo{volume}{411}},
  \bibinfo{pages}{839} (\bibinfo{year}{1994}),
  \href{http://arxiv.org/abs/hep-lat/9310022}{ArXiv:hep-lat/9310022}.

\bibitem[{\citenamefont{Bornyakov
  et~al.}(2010{\natexlab{a}})\citenamefont{Bornyakov, Horsley, Nakamura,
  Polikarpov, Rakow, and Schierholz}}]{Bornyakov:2011}
\bibinfo{author}{\bibfnamefont{V.~G.} \bibnamefont{Bornyakov}},
  \bibinfo{author}{\bibfnamefont{R.}~\bibnamefont{Horsley}},
  \bibinfo{author}{\bibfnamefont{Y.}~\bibnamefont{Nakamura}},
  \bibinfo{author}{\bibfnamefont{M.~I.} \bibnamefont{Polikarpov}},
  \bibinfo{author}{\bibfnamefont{P.}~\bibnamefont{Rakow}}, \bibnamefont{and}
  \bibinfo{author}{\bibfnamefont{G.}~\bibnamefont{Schierholz}},
  \bibinfo{journal}{PoS} \textbf{\bibinfo{volume}{2010}}, \bibinfo{pages}{170}
  (\bibinfo{year}{2010}{\natexlab{a}}),
  \href{http://arxiv.org/abs/1102.4461}{ArXiv:1102.4461}.

\bibitem[{\citenamefont{Bornyakov
  et~al.}(2010{\natexlab{b}})\citenamefont{Bornyakov, Horsley, Morozov,
  Nakamura, Polikarpov, Rakow, Schierholz, and Suzuki}}]{Bornyakov:2009}
\bibinfo{author}{\bibfnamefont{V.~G.} \bibnamefont{Bornyakov}},
  \bibinfo{author}{\bibfnamefont{R.}~\bibnamefont{Horsley}},
  \bibinfo{author}{\bibfnamefont{S.~M.} \bibnamefont{Morozov}},
  \bibinfo{author}{\bibfnamefont{Y.}~\bibnamefont{Nakamura}},
  \bibinfo{author}{\bibfnamefont{M.~I.} \bibnamefont{Polikarpov}},
  \bibinfo{author}{\bibfnamefont{P.~E.~L.} \bibnamefont{Rakow}},
  \bibinfo{author}{\bibfnamefont{G.}~\bibnamefont{Schierholz}},
  \bibnamefont{and} \bibinfo{author}{\bibfnamefont{T.}~\bibnamefont{Suzuki}},
  \bibinfo{journal}{Phys. Rev. D} \textbf{\bibinfo{volume}{82}},
  \bibinfo{pages}{014504} (\bibinfo{year}{2010}{\natexlab{b}}),
  \href{http://arxiv.org/abs/0910.2392}{ArXiv:0910.2392}.

\bibitem[{\citenamefont{Aoki et~al.}(2006{\natexlab{a}})\citenamefont{Aoki,
  Endrodi, Fodor, Katz, and Szabo}}]{Fodor:06:1}
\bibinfo{author}{\bibfnamefont{Y.}~\bibnamefont{Aoki}},
  \bibinfo{author}{\bibfnamefont{G.}~\bibnamefont{Endrodi}},
  \bibinfo{author}{\bibfnamefont{Z.}~\bibnamefont{Fodor}},
  \bibinfo{author}{\bibfnamefont{S.~D.} \bibnamefont{Katz}}, \bibnamefont{and}
  \bibinfo{author}{\bibfnamefont{K.~K.} \bibnamefont{Szabo}},
  \bibinfo{journal}{Nature} \textbf{\bibinfo{volume}{443}}, \bibinfo{pages}{675
  } (\bibinfo{year}{2006}{\natexlab{a}}),
  \href{http://arxiv.org/abs/hep-lat/0611014}{ArXiv:hep-lat/0611014}.

\bibitem[{\citenamefont{Aoki et~al.}(2006{\natexlab{b}})\citenamefont{Aoki,
  Fodor, Katz, and Szabo}}]{Fodor:06:2}
\bibinfo{author}{\bibfnamefont{Y.}~\bibnamefont{Aoki}},
  \bibinfo{author}{\bibfnamefont{Z.}~\bibnamefont{Fodor}},
  \bibinfo{author}{\bibfnamefont{S.~D.} \bibnamefont{Katz}}, \bibnamefont{and}
  \bibinfo{author}{\bibfnamefont{K.~K.} \bibnamefont{Szabo}},
  \bibinfo{journal}{Phys. Lett. B} \textbf{\bibinfo{volume}{643}},
  \bibinfo{pages}{46 } (\bibinfo{year}{2006}{\natexlab{b}}),
  \href{http://arxiv.org/abs/hep-lat/0609068}{ArXiv:hep-lat/0609068}.

\bibitem[{\citenamefont{Mizher et~al.}(2010)\citenamefont{Mizher, Chernodub,
  and Fraga}}]{Chernodub:10:2}
\bibinfo{author}{\bibfnamefont{A.~J.} \bibnamefont{Mizher}},
  \bibinfo{author}{\bibfnamefont{M.~N.} \bibnamefont{Chernodub}},
  \bibnamefont{and} \bibinfo{author}{\bibfnamefont{E.~S.} \bibnamefont{Fraga}},
  \bibinfo{journal}{Phys. Rev. D} \textbf{\bibinfo{volume}{82}},
  \bibinfo{pages}{105016} (\bibinfo{year}{2010}),
  \href{http://arxiv.org/abs/1004.2712}{ArXiv:1004.2712}.

\bibitem[{\citenamefont{Omelyan et~al.}(2003)\citenamefont{Omelyan, Mrygloda,
  and Folk}}]{Omelyan:03:1}
\bibinfo{author}{\bibfnamefont{I.~P.} \bibnamefont{Omelyan}},
  \bibinfo{author}{\bibfnamefont{I.~M.} \bibnamefont{Mrygloda}},
  \bibnamefont{and} \bibinfo{author}{\bibfnamefont{R.}~\bibnamefont{Folk}},
  \href{http://dx.doi.org/10.1016/S0010-4655(02)00754-3}{  \bibinfo{journal}{Comp.Phys.Commun.} \textbf{\bibinfo{volume}{151}},
  \bibinfo{pages}{272} (\bibinfo{year}{2003})}.

\bibitem[{\citenamefont{Takaishi and {de Forcrand}}(2006)}]{Takaishi:05:1}
\bibinfo{author}{\bibfnamefont{T.}~\bibnamefont{Takaishi}} \bibnamefont{and}
  \bibinfo{author}{\bibfnamefont{P.}~\bibnamefont{{de Forcrand}}},
  \bibinfo{journal}{Phys. Rev. E} \textbf{\bibinfo{volume}{73}},
  \bibinfo{pages}{036706} (\bibinfo{year}{2006}),
  \href{http://arxiv.org/abs/hep-lat/0505020}{ArXiv:hep-lat/0505020}.

\bibitem[{\citenamefont{Cundy et~al.}(2011)\citenamefont{Cundy, Kennedy, and
  Sch\"{a}fer}}]{Cundy:11:1}
\bibinfo{author}{\bibfnamefont{N.}~\bibnamefont{Cundy}},
  \bibinfo{author}{\bibfnamefont{T.}~\bibnamefont{Kennedy}}, \bibnamefont{and}
  \bibinfo{author}{\bibfnamefont{A.}~\bibnamefont{Sch\"{a}fer}},
  \bibinfo{journal}{Nucl. Phys. B} \textbf{\bibinfo{volume}{845}},
  \bibinfo{pages}{30 } (\bibinfo{year}{2011}),
  \href{http://arxiv.org/abs/1010.5629}{ArXiv:1010.5629}.

\bibitem[{\citenamefont{Chandrasekharan}(1999)}]{Chandrasekharan:99:1}
\bibinfo{author}{\bibfnamefont{S.}~\bibnamefont{Chandrasekharan}},
  \bibinfo{journal}{Phys. Rev. D} \textbf{\bibinfo{volume}{60}},
  \bibinfo{pages}{074503} (\bibinfo{year}{1999}),
  \href{http://arxiv.org/abs/hep-lat/9805015}{ArXiv:hep-lat/9805015}.

\bibitem[{\citenamefont{Banks and Casher}(1980)}]{Banks:80:1}
\bibinfo{author}{\bibfnamefont{T.}~\bibnamefont{Banks}} \bibnamefont{and}
  \bibinfo{author}{\bibfnamefont{A.}~\bibnamefont{Casher}},
  \href{http://dx.doi.org/10.1016/0550-3213(80)90255-2}{  \bibinfo{journal}{Nucl. Phys. B} \textbf{\bibinfo{volume}{169}},
  \bibinfo{pages}{103 } (\bibinfo{year}{1980})}.

\bibitem[{\citenamefont{Landsteiner et~al.}(2012)\citenamefont{Landsteiner,
  Megias, and {Pena-Benitez}}}]{Landsteiner:12:1}
\bibinfo{author}{\bibfnamefont{K.}~\bibnamefont{Landsteiner}},
  \bibinfo{author}{\bibfnamefont{E.}~\bibnamefont{Megias}}, \bibnamefont{and}
  \bibinfo{author}{\bibfnamefont{F.}~\bibnamefont{{Pena-Benitez}}},
  \emph{\bibinfo{title}{Anomalous transport from {Kubo} formulae}},
  \bibinfo{howpublished}{in Lect. Notes Phys. "Strongly interacting matter in
  magnetic fields" (Springer), edited by D. Kharzeev, K. Landsteiner, A.
  Schmitt, H.-U. Yee} (\bibinfo{year}{2012}),
  \href{http://arxiv.org/abs/1207.5808}{ArXiv:1207.5808}.

\bibitem[{\citenamefont{Basar and Dunne}(2012)}]{Basar:12:1}
\bibinfo{author}{\bibfnamefont{G.}~\bibnamefont{Basar}} \bibnamefont{and}
  \bibinfo{author}{\bibfnamefont{G.~V.} \bibnamefont{Dunne}},
  \emph{\bibinfo{title}{The chiral magnetic effect and axial anomalies}},
  \bibinfo{howpublished}{in Lect. Notes Phys. "Strongly interacting matter in
  magnetic fields" (Springer), edited by D. Kharzeev, K. Landsteiner, A.
  Schmitt, H.-U. Yee} (\bibinfo{year}{2012}),
  \href{http://arxiv.org/abs/1207.4199}{ArXiv:1207.4199}.

\bibitem[{\citenamefont{Fukushima et~al.}(2008)\citenamefont{Fukushima,
  Kharzeev, and Warringa}}]{Kharzeev:08:2}
\bibinfo{author}{\bibfnamefont{K.}~\bibnamefont{Fukushima}},
  \bibinfo{author}{\bibfnamefont{D.~E.} \bibnamefont{Kharzeev}},
  \bibnamefont{and} \bibinfo{author}{\bibfnamefont{H.~J.}
  \bibnamefont{Warringa}}, \bibinfo{journal}{Phys. Rev. D}
  \textbf{\bibinfo{volume}{78}}, \bibinfo{pages}{074033}
  (\bibinfo{year}{2008}), \href{http://arxiv.org/abs/0808.3382}{ArXiv:0808.3382}.

\bibitem[{\citenamefont{Kharzeev
  et~al.}(2008{\natexlab{b}})\citenamefont{Kharzeev, McLerran, and
  Warringa}}]{Kharzeev:08}
\bibinfo{author}{\bibfnamefont{D.~E.} \bibnamefont{Kharzeev}},
  \bibinfo{author}{\bibfnamefont{L.~D.} \bibnamefont{McLerran}},
  \bibnamefont{and} \bibinfo{author}{\bibfnamefont{H.~J.}
  \bibnamefont{Warringa}}, \bibinfo{journal}{Nucl. Phys. A}
  \textbf{\bibinfo{volume}{803}}, \bibinfo{pages}{227}
  (\bibinfo{year}{2008}{\natexlab{b}}),
  \href{http://arxiv.org/abs/0711.0950}{ArXiv:0711.0950}.

\end{thebibliography}

\end{document}